\documentclass[journal,comsoc]{IEEEtran}
\usepackage[compatibility=false]{caption}
\usepackage{cite} 
\usepackage{booktabs} 
\usepackage{array}
\usepackage{tabularx} 
\usepackage{booktabs}
\usepackage{graphicx} 
\usepackage{caption} 
\usepackage{multirow} 
\usepackage{xcolor} 
\usepackage{amsthm}
\usepackage{subcaption}
\usepackage{amsmath,amsfonts}
\usepackage{textcomp}
\usepackage{makecell} 
\usepackage{helvet}
\usepackage{stfloats}
\usepackage{verbatim}
\usepackage{subscript} 
\usepackage[caption=false,font=normalsize,labelfont=sf,textfont=sf]{subfig}
\usepackage{algorithmic}
\usepackage[utf8]{inputenc}
\usepackage{amssymb}
\usepackage{amsthm}
\usepackage{cancel}  

\begin{document}
\setlength{\columnsep}{0.245 in}

\title{A Subcarrier-Aware Approach for Robust Respiratory Monitoring with Commodity Wi-Fi}

\author{Pei Tang, Yunpeng Ge
        and Ivan Wang-Hei Ho, \IEEEmembership{Senior Member, IEEE}
\thanks{This work was supported in part by the Smart Traffic Fund (Project No. PSRI/31/2202/PR) established under the Transport Department of the Hong Kong Special Administrative Region (HKSAR), China.}
\thanks{P. Tang, Y. Ge, and I. W.-H. Ho are with the Department of Electrical and Electronic Engineering, The Hong Kong Polytechnic University, Hong Kong, China. (e-mail: \{pei.tang, yunpeng.ge\}@connect.polyu.hk; ivanwh.ho@polyu.edu.hk)}}

\maketitle
\thispagestyle{empty}

\begin{abstract}
Wi-Fi sensing has emerged as a promising modality for contact-free respiratory monitoring in home healthcare due to its ubiquity. However, conventional approaches typically treat Channel State Information (CSI) subcarriers uniformly, neglecting their heterogeneous responses to breathing motions. To address this limitation, we propose a subcarrier-aware sensing framework that characterizes subcarrier-dependent respiratory sensitivities and exploits them for informative subcarrier selection. Based on the statistical properties of breathing-sensitive subcarriers, we further develop an unsupervised clustering-based breathing estimation method for robust respiratory monitoring. Experimental results show that the proposed method achieves 97\% breathing rate estimation accuracy and reduces the mean absolute error (MAE) by 0.45 breaths per minute (bpm) compared with baseline methods. Furthermore, we extend the analysis to breathing-pause detection, which derives an amplitude-attenuation-based threshold mechanism. Under breathing pauses, the proposed method achieves an MAE of 1.6 bpm in breathing rate estimation. These results demonstrate the robustness and effectiveness of the proposed framework across different scenarios and its potential for contactless home healthcare monitoring.

\end{abstract}


\section{Introduction}

\IEEEPARstart{H}{uman} respiratory monitoring is essential for evaluating respiratory function \cite{wakili_2025_smarthealth,xie_2023_tbcas}, detecting early signs of deterioration \cite{toften_2024_contactless,salem_2021_markov}, and managing chronic diseases \cite{larssen_2025_jmir,fan_2024_pattern}. As populations age and solitary living increases, continuous, long-term monitoring has become essential for the early detection of health decline \cite{wang_2024_OmniResMonitor,Alzaabi_2024_eldercare}. Traditional contact-based methods, such as nasal airflow sensors and thoracic belts \cite{naik_2023_biosensors,collop_2007_clinic}, remain the clinical gold standard due to their high accuracy. However, their obtrusiveness, need for professional attachment, and physical tethers make them unsuitable for comfortable, long-term, unsupervised use in home environments \cite{peng_2024_ieeetransinstrummeas,wakili_2025_survey}. This fundamental limitation has spurred the search for non-invasive alternatives that can seamlessly integrate into daily life.

The rapid proliferation of the Internet of Things (IoT) has made contactless sensing a promising solution for home-based health monitoring \cite{abu_ali_2024_survey,guan_2024_activity}. Among these technologies, Wi-Fi-based sensing has attracted significant attention for its unique combination of ubiquity \cite{he_2026_placement}, cost-effectiveness \cite{wang_2024_SlpRoF}, and minimal intrusiveness \cite{tang_2024_bicsi}. By exploiting the fine-grained Channel State Information (CSI) available from commercial Wi-Fi devices, researchers have demonstrated the capability to detect subtle physiological movements, such as chest displacements caused by respiration \cite{zhang_2023_nlos,xie_2024_interfering,shi_2023_ieeetransinstrummeas}.

Existing CSI-based sensing approaches typically partition the wireless channel into static and dynamic paths~\cite{wang_2016_model,he_2026_placement}, assuming that all subcarriers experience identical modulation patterns under target motion. However, recent studies have shown that subcarrier selection can significantly improve sensing performance~\cite{ge_2025_WiRe,pu_2026_selection}, suggesting that different subcarriers exhibit heterogeneous responses to the same physical movement. Such heterogeneity mainly arises from frequency-selective multipath propagation, where some subcarriers are more sensitive to target motion while others are dominated by noise or interference~\cite{Kanda_selection}. Moreover, this heterogeneity is further influenced by environmental conditions~\cite{he_2026_placement}, hardware characteristics~\cite{kong_2024_CSI-RFF}, and sensing tasks~\cite{ge_2022_review}. Consequently, existing approaches that treat all subcarriers uniformly fail to fully exploit subcarrier-dependent sensing information, which limits robustness and accuracy in practical respiratory monitoring scenarios.

Within the context of respiratory monitoring, an additional challenge arises from abnormal breathing patterns, particularly breathing pauses associated with sleep apnea syndrome~\cite{collop_2007_clinic,ambalkar_2023_infocom}. During such pauses, the respiratory signal amplitude drops abruptly to near zero, violating the continuous sinusoidal assumption underlying existing respiration models~\cite{george_2005_sinusoidal,zhang_2023_nlos}. Although prior studies have explored pause detection for apnea monitoring~\cite{ge_2025_WiRe,liu_respiration}, the theoretical impact of respiratory interruptions on breathing estimation remains insufficiently characterized. Therefore, developing robust mechanisms to handle respiratory discontinuities is important for practical contact-free respiratory monitoring.

To address these limitations, this paper introduces a subcarrier-aware sensing framework for Wi-Fi respiratory monitoring. We first develop a subcarrier-dependent model to characterize heterogeneous respiratory sensitivities across CSI subcarriers and design an unsupervised clustering method for sensitive subcarrier selection. Furthermore, we extend the framework to respiratory pause scenarios and develop a robust pause-aware breathing sensing mechanism. These contributions advance contact-free respiratory monitoring toward practical IoT-enabled healthcare applications. The main contributions of this paper are threefold:
\begin{enumerate}
    \item \textbf{Subcarrier-aware sensing model:} 
   We propose a subcarrier-dependent sensing model that accounts for the heterogeneous respiratory sensitivities of CSI subcarriers. Based on this model, we further analyze the statistical feature relationships of sensitive subcarriers, which provides theoretical support for subcarrier-aware respiratory sensing.

     \item \textbf{Subcarrier-selective breathing estimation algorithm:} 
    Based on the proposed model, we develop an unsupervised subcarrier selection and breathing estimation framework for robust respiratory monitoring. Experimental results demonstrate 97\% breathing rate estimation accuracy and reduce the mean absolute error (MAE) by 0.45 breaths per minute (bpm) compared with baseline methods.

    \item \textbf{Robust breathing pause detection:} 
    We further extend the framework to respiratory pause scenarios and develop a threshold-based detection mechanism. Under pause conditions, our method achieves an MAE of 1.6 bpm and improves accuracy by 9\% over baselines.

\end{enumerate}

The remainder of this paper is organized as follows. Section II reviews related work. Section III presents the theoretical foundations of the proposed algorithm, while Section IV details the framework. Section V discusses the experimental results and Section VI concludes the paper.

\section{Related Work}

\subsection{Contactless Respiratory Monitoring}
Human respiration induces subtle chest movements that can modulate various physical fields, including electromagnetic waves \cite{xie_2023_tbcas}, acoustic signals \cite{wang_2024_OmniResMonitor}, and light \cite{khanam_2021_light}, thereby enabling contactless sensing. A range of technologies have been explored for this purpose. Among these, radar-based systems offer high sensitivity and range resolution but require dedicated hardware, limiting their deployment in daily living environments \cite{wang_2025_radar}. Optical approaches, including infrared and video cameras, can capture breathing through thermal or motion imagery, yet they raise privacy concerns and are sensitive to lighting conditions \cite{makkapati_2016_camera}. Acoustic sensing leverages speakers and microphones on smart devices, but its performance can be degraded by ambient noise \cite{wakili_2025_survey}.

Among these alternatives, Wi-Fi-based sensing has emerged as a promising modality due to its ubiquity, cost-effectiveness, and non-intrusive nature \cite{abu_ali_2024_survey}. Early Wi-Fi sensing efforts relied on received signal strength indicator (RSSI), which measures the average power of the received signal \cite{abdelnasser_2015_RSS}. While RSSI is readily available on Wi-Fi devices, its coarse granularity and susceptibility to multipath effects limit its ability to capture fine-grained physiological motions \cite{liu_respiration}. By contrast, CSI provides fine-grained amplitude and phase measurements for multiple orthogonal frequency-division multiplexing (OFDM) subcarriers, enabling the detection of movements such as chest displacements during breathing \cite{song_2025_finersense,zeng_2019_Farsense}. This rich, multi-dimensional representation forms the physical basis for the work presented in this paper.

\subsection{CSI-Based Breathing Sensing Model}
In a typical Wi-Fi system, the relationship between transmitted and received signals can be modeled as \cite{zhang_2023_nlos}:
\begin{equation}
\mathbf{y}(t) = \mathbf{H}(t) \mathbf{x}(t) + \mathbf{n}(t)
\end{equation}
where $\mathbf{x}(t) \in \mathbb{C}^{N_T}$ and $\mathbf{y}(t) \in \mathbb{C}^{N_R}$ are the transmitted and received signal vectors at time $t$, respectively, with $N_T$ and $N_R$ denoting the number of transmit and receive antennas. The matrix $\mathbf{H}(t) \in \mathbb{C}^{N_R \times N_T}$ represents the channel state information, which captures the amplitude and phase responses of the wireless channel. The term $\mathbf{n}(t)$ accounts for additive noise.

For each transmit-receive antenna pair, modern Wi-Fi systems employ OFDM, partitioning the channel into $N$ orthogonal subcarriers. The CSI at time $t$ can thus be represented as a vector \cite{tang_2024_bicsi}:
\begin{equation}
\mathbf{H}(t) = [H_1(t), H_2(t), \dots, H_N(t)]^T
\end{equation}
where $H_n(t) = |H_n(t)| e^{j \angle H_n(t)}$ denotes the complex channel response for subcarrier $n$, with $|H_n(t)|$ and $\angle H_n(t)$ representing its amplitude and phase, respectively.

To model the effect of breathing on the wireless channel, prior work has traditionally adopted a time-varying multipath model. The channel response for a given frequency $f$ can be expressed as the superposition of static and dynamic components \cite{wang_2016_model,xie_2024_interfering}:
\begin{equation}
\label{eq:traditional sensing model}
H(f, t) = H_s(f) + \sum_{p=1}^P H_{d,k}(f, t) + \epsilon(f, t)
\end{equation}
where $H_s(f)$ represents static contributions, $H_{d,k}(f, t)$ denotes the contribution of the $p$-th dynamic path, and $\epsilon(f, t)$ is noise. In the context of respiration monitoring, the dominant dynamic path is typically attributed to chest displacement during breathing. This motion is commonly approximated as a single sinusoidal oscillation \cite{xie_2024_interfering}:

\begin{equation}
H_d(f, t) = A e^{-j \frac{2\pi}{\lambda_f} d(t)}, \quad d(t) = d_0 + \Delta d \sin(\omega t).
\end{equation}
where $d(t)$ is the time-varying chest displacement, $d_0$ is the mean distance, $\Delta d$ is the displacement amplitude, $\omega$ is the angular breathing frequency, and $\lambda_f = c/f$ is the wavelength.

\begin{figure*}[ht]  
    \centering
    \begin{minipage}[b]{0.32\linewidth} 
        \centering
        \includegraphics[width=\linewidth]{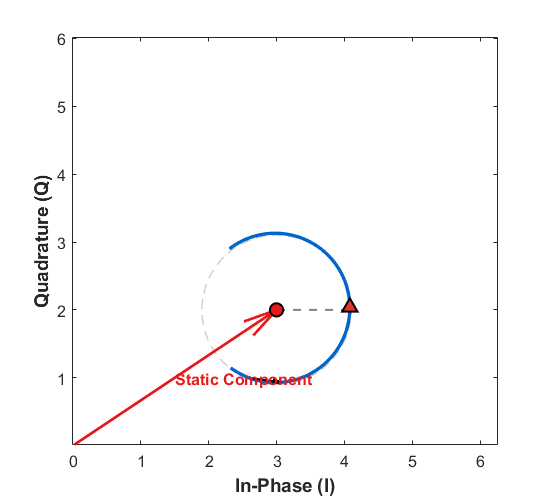} 
        \caption{Theoretical model representation in the complex plane.}
        \label{fig:1}
    \end{minipage}%
    \hfill
    \begin{minipage}[b]{0.32\linewidth} 
        \centering
        \includegraphics[width=\linewidth]{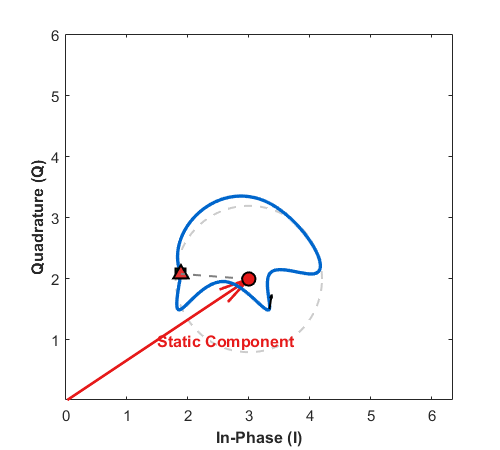}  
        \caption{The actual trajectory of one subcarrier.}
        \label{fig:2}
    \end{minipage}%
    \hfill
    \begin{minipage}[b]{0.32\linewidth}  
        \centering
        \includegraphics[width=\linewidth]{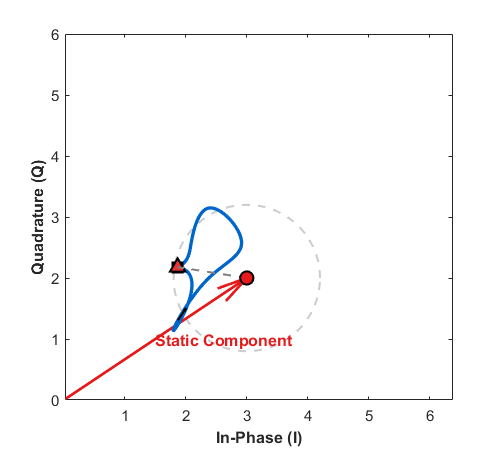} 
        \caption{The actual trajectory of another subcarrier.}
        \label{fig:3}
    \end{minipage}
\end{figure*}

This formulation carries a critical implicit assumption: all subcarriers experience the same breathing-induced modulation pattern, despite their frequency-dependent propagation characteristics. Although OFDM provides measurements across multiple frequency-distinct subcarriers, the model suggests that each subcarrier's complex trajectory should trace a perfect circle \cite{zeng_2019_Farsense,zhang_2023_nlos}, as illustrated in Fig.~\ref{fig:1}. In realistic indoor environments, however, each subcarrier's frequency-dependent propagation leads to a unique superposition of multipath components, challenging this idealized view.

\subsection{Advances in Wi-Fi Breathing Monitoring}
Building on the above sensing model, existing research has investigated CSI-based respiratory monitoring using signal processing and machine learning approaches~\cite{ge_2022_review,abu_ali_2024_survey}. Early studies mainly relied on signal-processing techniques, including Fast Fourier transform (FFT)-based spectral analysis~\cite{liu_2021_FFT}, signal filtering~\cite{Ge_2024_RMS}, and Fresnel-zone-based propagation modeling~\cite{wang_2016_model,xie_2023_tbcas}, to extract or characterize respiration-induced CSI fluctuations. To improve sensing robustness, subsequent studies further combined CSI amplitude and phase information to exploit their complementary sensitivities to respiratory motion~\cite{li_2022_Diversense, long_2022_Adaptive}.

More recently, machine learning approaches have been introduced to address complex respiratory scenarios where conventional signal-processing methods become less effective~\cite{wakili_2025_survey, ambalkar_2023_infocom}. These methods are mainly applied to breathing pattern recognition and abnormal respiration analysis~\cite{fan_2024_pattern,zhang_2024_healthcare}. Although these approaches have demonstrated promising performance, they generally require extensive labeled datasets and often exhibit limited generalization across different environments and subjects~\cite{wakili_2025_smarthealth,larssen_2025_jmir}.

Despite these advances, the heterogeneous respiratory sensitivities of CSI subcarriers remain insufficiently characterized. As discussed in Section II.B, conventional sensing models assume that all subcarriers follow identical circular trajectories under respiratory motion. However, practical CSI trajectories significantly deviate from this assumption. As illustrated in Figs. 2-3, different subcarriers exhibit distinct non-circular responses to breathing motions, revealing substantial subcarrier heterogeneity in respiration sensing. Several subcarrier-selection strategies have therefore been proposed to improve sensing performance, such as WiRe based on signal variance [22] and RMS based on breathing energy [37]. Although these methods demonstrate the effectiveness of subcarrier selection, the underlying characteristics of subcarrier heterogeneity and its adaptive exploitation remain insufficiently explored.

Another underexplored challenge arises from abnormal respiratory conditions, particularly breathing pauses. Existing studies mainly address pause detection using data-driven classification methods~\cite{wakili_2025_survey,ge_2022_review}, while the underlying CSI dynamics during respiratory interruptions remain insufficiently understood. Recent observations indicate that CSI signals exhibit distinctive fluctuations during breathing pauses~\cite{ge_2025_WiRe,liu_respiration}. However, abrupt respiratory interruptions violate the continuous sinusoidal assumption adopted in conventional respiration models, potentially introducing biases into respiratory rate estimation~\cite{collop_2007_clinic,abu_ali_2024_survey}. Therefore, modeling respiratory discontinuities is important for robust Wi-Fi-based respiratory monitoring and apnea assessment.

Motivated by these limitations, we develop a subcarrier-aware sensing framework that characterizes the heterogeneous respiratory sensitivities of CSI subcarriers and adaptively exploits such heterogeneity for robust respiratory sensing. Furthermore, we extend the framework to respiratory pause scenarios, thereby advancing Wi-Fi-based respiratory monitoring in both theory and practice.

\begin{figure*}[ht]
    \centering
    \includegraphics[width=0.8\linewidth]{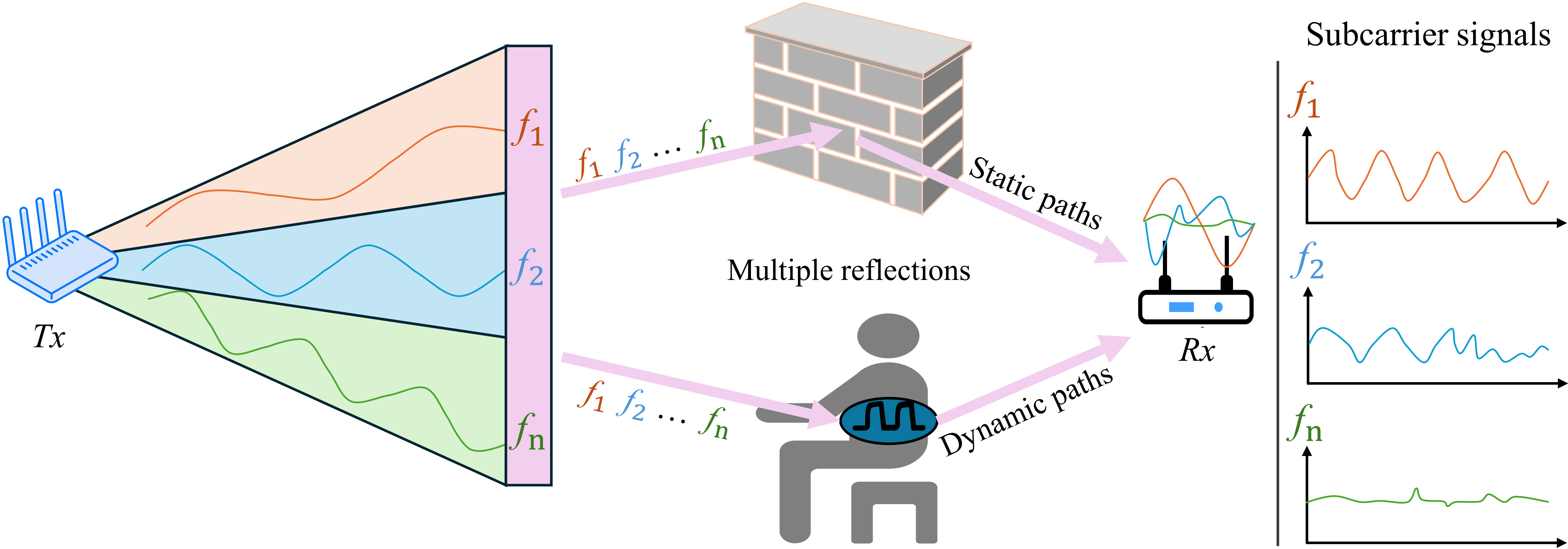}
    \caption{The illustration of subcarrier-dependent breathing sensitivity.}
    \label{fig:sensing model}
\end{figure*}

\section{Mathematical Model}

\subsection{Subcarrier-Dependent CSI Model for Breathing Sensing}

As discussed in Section II.B, conventional CSI sensing models implicitly assume that all subcarriers experience identical respiratory modulation patterns. However, as illustrated in Fig.~\ref{fig:sensing model}, due to dynamic multipath reflections, the received CSI time series across different subcarriers exhibit varying sensitivities to breathing-induced chest movements.

Building upon the traditional sensing model in (\ref{eq:traditional sensing model}), we further consider the subcarrier-dependent respiratory responses caused by multipath superposition and frequency-selective propagation. To capture this heterogeneity, the CSI response on subcarrier $n$ can be expressed as:

\begin{equation}
H_n(t)=H_{s,n}+\gamma_n e^{-j2\pi f_n d(t)/c}+\epsilon_n(t),
\label{eq}
\end{equation}
where $H_{s,n}$ denotes the static component on subcarrier $n$, $\gamma_n$ is an effective respiratory sensitivity coefficient, $f_n$ is the center frequency, $c$ is the speed of light, and $\epsilon_n(t)$ represents measurement noise.

Because each subcarrier operates at a different frequency, its propagation response to respiration-induced multipath variations may differ. Consequently, multipath superposition and frequency-selective propagation lead to heterogeneous respiratory sensitivities across subcarriers. As a result, some subcarriers exhibit stronger respiration-induced fluctuations, while others respond only weakly. This heterogeneity is also evident after group-wise normalization, where some subcarriers exhibit substantially stronger breathing signatures than others, as shown in Fig.~\ref{fig:select1}.

Accordingly, the collection of all subcarrier responses forms the set:
\[
\mathcal{H} = \{\mathbf{H}_1, \mathbf{H}_2, \dots, \mathbf{H}_n\} = \mathcal{H}^{\text{sens}} \cup \mathcal{H}^{\text{insens}}
\]
where $\mathcal{H}^{\text{sens}}$ and $\mathcal{H}^{\text{insens}}$ contain breathing-sensitive and breathing-insensitive subcarrier responses, respectively.

\subsubsection{Breathing Signal Characteristics in Time Domain}
For a subcarrier whose response belongs to $\mathcal{H}^{\text{sens}}$, the CSI signal exhibits periodic modulation due to chest movement \cite{george_2005_sinusoidal}. This can be modeled in the continuous time domain as:
\begin{equation}
x_n^{\text{sens}}(t) = A_n + B_n \sin(2\pi r_b t + \phi_n) + \epsilon_n(t), \quad B_n > 0.
\label{eq:continuous_model}
\end{equation}
where:
\begin{itemize}
\item $A_n$: Subcarrier intrinsic baseline determined by static propagation paths,
\item $B_n$: Breathing modulation amplitude,
\item $r_b \in [0.1, 0.5]$ Hz: Breathing frequency,
\item $\phi_n$: Phase offset,
\item $\epsilon_n(t)$: Time-variant noise.
\end{itemize}

Conversely, for a subcarrier response in $\mathcal{H}^{\text{insens}}$, the signal lacks such periodicity and remains approximately constant:
\begin{equation}
x_n^{\text{insens}}(t) = C_n + \delta_n(t)
\end{equation}
where $C_n$ represents the constant amplitude from static paths and $\delta_n(t)$ captures environmental noise. 

Overall, the modulation amplitude $B_n$ varies significantly across subcarriers. Subcarriers with large $B_n$ (i.e., $\mathbf{H}_n \in \mathcal{H}^{\text{sens}}$) carry more obvious breathing signature, while those with $B_n \approx 0$ (i.e., $\mathbf{H}_n \in \mathcal{H}^{\text{insens}}$) do not. This subcarrier-dependent behavior motivates our proposed subcarrier-aware mechanism. To this end, the next section introduces a group-wise normalization procedure.

\subsection{Group-wise Normalization Theory}
To facilitate sensitive subcarrier selection, we propose a group-wise normalization scheme across subcarriers within each time group. It preserves periodic patterns from sensitive subcarriers, suppresses those from insensitive subcarriers, and eliminates the group-dependent baseline.

Specifically, we partition the time series into non-overlapping groups (e.g., one per second). The one-second window is selected based on the expected breathing frequency range and is used as a smoothing step before group-wise normalization. Within each group m, we compute the average CSI magnitude across its $L$ samples, which forms a group-average vector:

\begin{equation}
x_{m,n} = \frac{1}{L}\sum_{t \in \mathcal{T}_m} x_n(t), \quad \mathcal{T}_m \text{ is the time interval of group } m.
\end{equation}
Applying the continuous-time model in ~\eqref{eq:continuous_model} to this averaging process yields the discrete group-average model:
\begin{equation}
x_{m,n} = s_{m,n} + b_m + \epsilon_{m,n}
\end{equation}
where $s_{m,n}$: Breathing-induced signal component, group-dependent baseline shared by subcarriers within the group, and $\epsilon_{m,n}$: Residual measurement noise.

\begin{figure*}[ht]  
    \centering
    \begin{minipage}[b]{0.32\linewidth} 
        \centering
        \includegraphics[width=\linewidth]{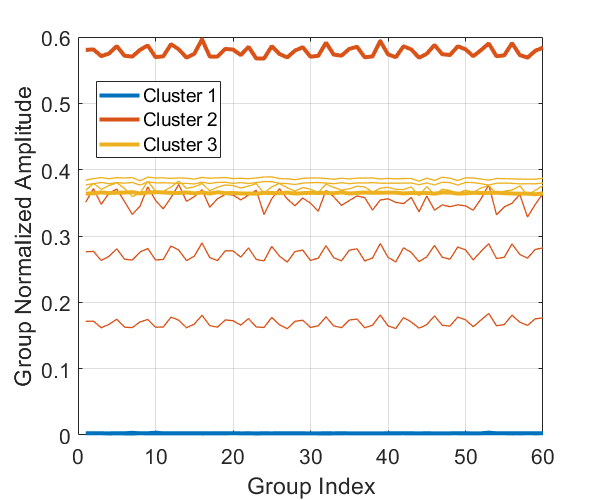} 
        \caption{Illustration of subcarriers after group-wise normalization.}
        \label{fig:select1}
    \end{minipage}%
    \hfill
    \begin{minipage}[b]{0.32\linewidth} 
        \centering
        \includegraphics[width=\linewidth]{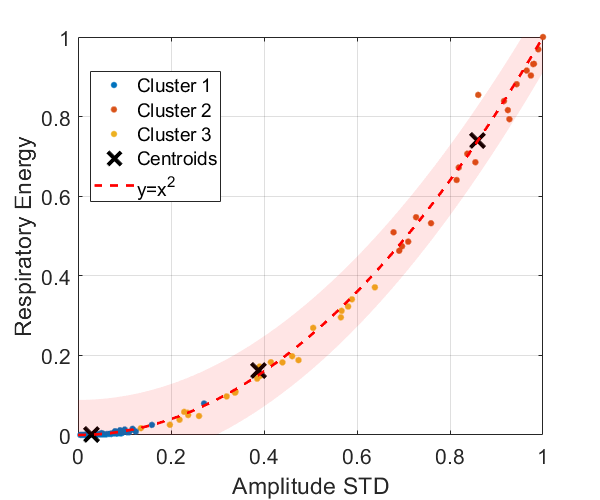}  
        \caption{The relationship between breath energy and amplitude STD.}
        \label{fig:select2}
    \end{minipage}%
    \hfill
    \begin{minipage}[b]{0.32\linewidth}  
        \centering
        \includegraphics[width=\linewidth]{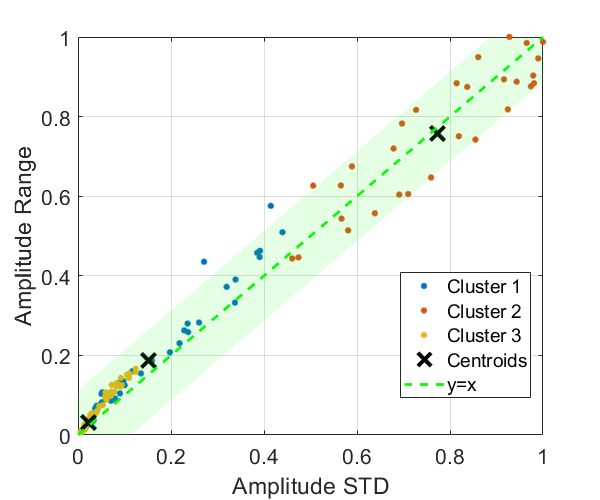} 
        \caption{The relationship between amplitude range and amplitude STD.}
        \label{fig:select3}
    \end{minipage}
\end{figure*}

The group-wise normalization transforms $\mathbf{x}_m$ to $\mathbf{y}_m$:

\begin{align}
y_{m,n} = \frac{x_{m,n} - \alpha_m}{\beta_m} &= \frac{s_{m,n} + \epsilon_{m,n} - \min_{j} (s_{m,j} + \epsilon_{m,j})}{\beta_m}\\
\alpha_m &= \min_{n} x_{m,n}  \\
\beta_m &= \max_{n} x_{m,n} - \min_{n} x_{m,n}
\end{align}

The effects of group-wise normalization include:

\subsubsection{Proposition 1 (Preservation of Breathing Periodicity)} For breathing-sensitive subcarriers, group-wise normalization preserves the sinusoidal breathing pattern.

\begin{proof}
Consider a sensitive subcarrier $s$ with $x_{m,s} = A_s + B_s \sin(\theta_m) + \epsilon_{m,s}$, where $\theta_m = 2\pi r_b t_m + \phi_s$.

After normalization:

\begin{equation}
\begin{aligned}
y_{m,s} &= \frac{A_s + B_s \sin(\theta_m) + \epsilon_{m,s} - \alpha_m}{\beta_m} \\
&= \frac{A_s - \alpha_m}{\beta_m} + \frac{B_s}{\beta_m} \cdot \underbrace{\sin(\theta_m)}_{\substack{\text{breathing pattern}}} + \frac{\epsilon_{m,s}}{\beta_m}
\end{aligned}
\end{equation}

The breathing periodicity is preserved because the sinusoidal structure of $\sin(\theta_m)$ remains unchanged.
\end{proof}

\subsubsection{Proposition 2 (Compression to Near-Constant Values)}
For breathing-insensitive subcarriers, group-wise normalization compresses their values to near-constant levels across groups.

\begin{proof}
For an insensitive subcarrier $i$ with $x_{m,i} = C_i + \delta_{m,i}$, where $|\delta_{m,i}| \ll C_i$:

\begin{align}
y_{m,i} &= \frac{C_i + \delta_{m,i} - \alpha_m}{\beta_m} \\
&= \underbrace{\frac{C_i - \alpha_m}{\beta_m}}_{k_i} + \frac{\delta_{m,i}}{\beta_m}
\end{align}

As commonly assumed [6,21], the noise term $\delta_{m,i}$ is typically smaller than $\beta_m$, so the ratio $\delta_{m,i}/\beta_m$ becomes negligible. Therefore, $y_{m,i} \approx k_i$ across different groups $m$.
\end{proof}

Therefore, after group-wise normalization, sensitive subcarriers appear as fluctuating curves, while insensitive subcarriers appear as near-straight lines in group-index plots.

\subsubsection{Proposition 3 (Baseline removal)} The group-dependent baseline $b_m$ is eliminated through group-wise normalization.

\begin{proof}
Please refer to Appendix A.1 for the proof of Proposition 3.
\end{proof}

Moreover, the effect of group-wise normalization on different subcarrier types is illustrated in Fig.~\ref{fig:select1}. For visualization purposes, we pre-classify subcarriers into three groups based on their breathing sensitivity (detailed clustering method in Section IV-B). As shown, highly sensitive subcarriers (cluster 2) retain clear sinusoidal patterns after normalization, while less sensitive ones (clusters 1 and 3) are compressed to near-constant levels. This demonstrates the effectiveness of group-wise normalization in distinguishing informative subcarriers for subsequent processing.

\subsection{Statistical Properties for Breathing Subcarrier Selection}

For breathing-sensitive subcarriers with significant periodic breathing modulation, the normalized signal $y_n(m)$ can be modeled as:

\begin{equation}
y_n(m) = \mu_n + \kappa_n \sin(2\pi r_b t_m + \phi_n) + \zeta_n(m)
\end{equation}
where $\mu_n$ is the mean value after group-wise normalization; $\kappa_n$ is the normalized breathing modulation amplitude; $t_m$ is the time instant for group $m$; and $\zeta_n(m)$ is the residual noise after normalization.

For each subcarrier $n$, we extract three key features from its amplitude time series $y_n(m)$:
\begin{itemize}
\item Standard deviation: $\sigma_n = \sqrt{\frac{1}{M}\sum_{m=1}^M (y_n(m) - \bar{y}_n)^2}$, where $M$ is the number of samples.
\item Amplitude range: $R_n = \max_m y_n(m) - \min_m y_n(m)$
\item Breathing-band energy: $E_n = \sum_{f \in \mathcal{F}_b} P_n(f)$, where $P_n(f)$ is the power spectral density of $y_n(m)$, and $\mathcal{F}_b = [0.1, 0.5]$ Hz is the typical breathing frequency range adopted in prior works \cite{ge_2025_WiRe,Alzaabi_2024_eldercare}.
\end{itemize}

Then we can derive the following properties after the group-wise normalization:

\textit{Proposition 4 (Normalized Feature Relationships):} 
For breathing-sensitive subcarriers with dominant sinusoidal breathing modulation of amplitude $\kappa_n$, the group-wise normalized features satisfy:
\begin{align}
\tilde{R}_n &\propto \tilde{\sigma}_n \label{eq:norm_sigma_range} \\
\tilde{E}_n &\propto (\tilde{\sigma}_n)^2 \label{eq:norm_sigma_energy}
\end{align}
where $\tilde{\sigma}_n$, $\tilde{R}_n$, and $\tilde{E}_n$ are the min-max normalized versions of $\sigma_n$ (STD), $R_n$, and $E_n$ respectively.

\begin{proof}
Please refer to Appendix B.1 for the proof of Proposition 4.
\end{proof}

These theoretical relationships reveal a fundamental property of the normalized feature space for subcarriers with different sensitivities. As illustrated in Figs.~\ref{fig:select2}--\ref{fig:select3}, subcarriers with stronger breathing modulation exhibit structured patterns along the parabola $\tilde{E}_n \propto (\tilde{\sigma}_n)^2$ and the line $\tilde{R}_n \propto \tilde{\sigma}_n$, with these highly sensitive subcarriers (cluster 2) concentrating in the region of larger feature values. In contrast, less informative subcarriers (clusters 1 and 3) tend to cluster near the origin. These observations, consistent with our theoretical analysis, provide the basis for using unsupervised clustering to identify sensitive subcarriers in Section IV-B.

\subsection{Information Fusion of Amplitude and Phase}

\begin{figure}[t]
\centering
    \includegraphics[width=0.8\linewidth]{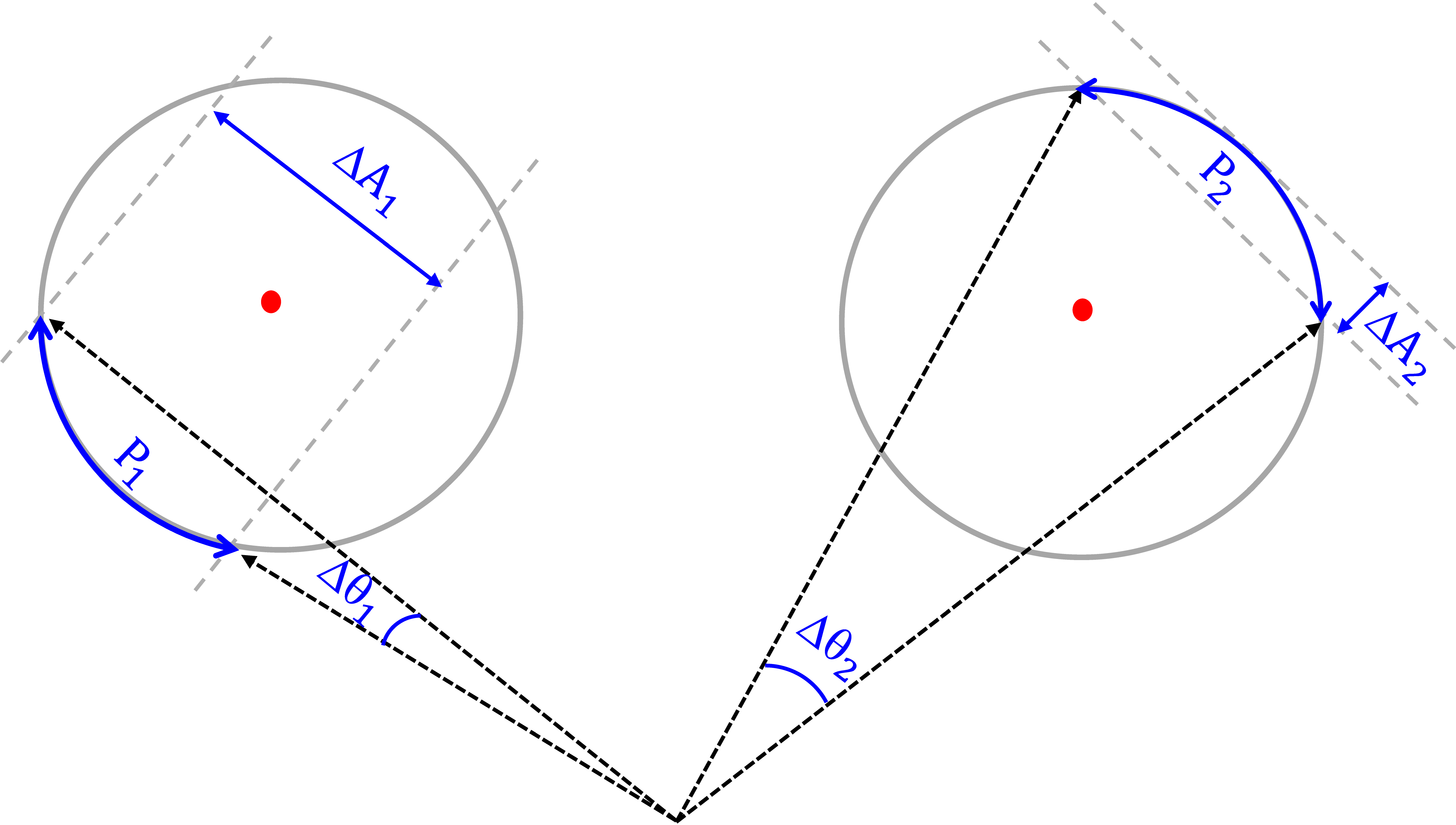}
    \caption{The illustration of information fusion.}
    \label{fig:fusion}
\end{figure} 

In addition to subcarrier selection, the proposed algorithm fuses CSI amplitude and phase information to leverage their complementary sensitivity to breathing motions. As illustrated in Fig.~\ref{fig:fusion}, the relative sensitivity of amplitude and phase varies across different propagation paths. For path $P_1$, amplitude variation $\Delta A_1$ dominates over phase variation $\Delta\theta_1$, making amplitude more informative. Conversely, for path $P_2$, phase variation $\Delta\theta_2$ exceeds amplitude variation $\Delta A_2$, which renders the phase more sensitive. By adaptively selecting the modality with higher sensitivity (quantified by $R^2$), our algorithm achieves robust breathing rate estimation across diverse environments, as demonstrated in Section IV-C.

\begin{figure*}[hb]  
    \centering
    \begin{minipage}[b]{0.32\linewidth} 
        \centering
        \includegraphics[width=\linewidth]{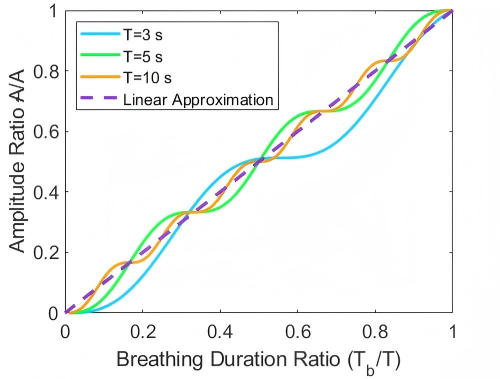} 
        \caption{Illustration of oscillatory behavior around linear estimation.}
        \label{fig:stop1}
    \end{minipage}%
    \hfill
    \begin{minipage}[b]{0.32\linewidth} 
        \centering
        \includegraphics[width=\linewidth]{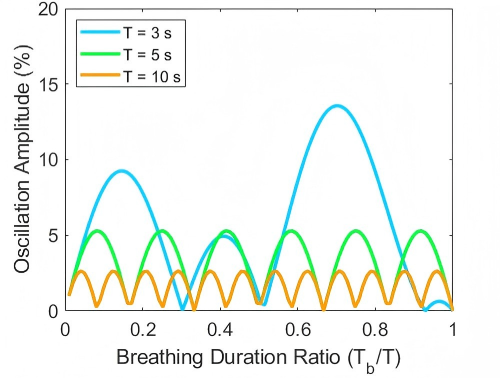}  
        \caption{Oscillatory behavior under different observation durations.}
        \label{fig:stop2}
    \end{minipage}%
    \hfill
    \begin{minipage}[b]{0.32\linewidth}  
        \centering
        \includegraphics[width=\linewidth]{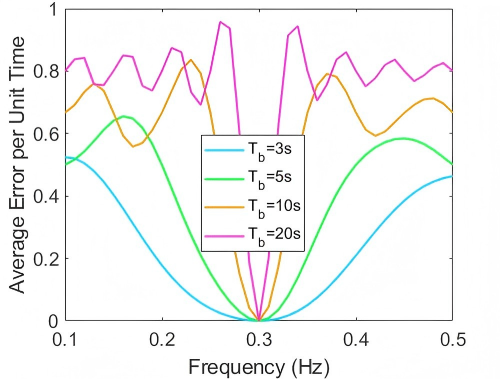} 
        \caption{Mean Error Under Different Observation durations}
        \label{fig:stop3}
    \end{minipage}
\end{figure*}

\subsection{Impact of Breathing Pauses on Sinusoidal Fitting}

\label{sec:theoretical-analysis}
This section analyzes how breathing pauses affect sinusoidal parameter estimation. Let $T$ denote the total observation time, with breathing occurring during $T_b$ seconds ($0 \leq T_b \leq T$) followed by a pause during $T_a = T - T_b$ seconds. The respiratory signal is modeled as:

\begin{equation}
    s(t) = 
    \begin{cases}
        A \sin(2\pi r_b t + \phi) + C, & 0 \leq t < T_b \\
        0, & T_b \leq t \leq T
    \end{cases}
    \label{eq:signal_model}
\end{equation}
where $A$, $r_b$, $\phi$, and $C$ represent the true amplitude, breathing frequency, phase, and baseline offset, respectively.

We attempt to fit the entire signal $s(t)$ over $[0, T]$ using a sinusoidal model:

\begin{equation}
    \hat{s}(t) = \hat{A} \sin(2\pi \hat{r_b} t + \hat{\phi}) + \hat{C}
    \label{eq:fit_model}
\end{equation}
where $\hat{A}$, $\hat{r_b}$, $\hat{\phi}$, and $\hat{C}$ are the estimated parameters obtained through least-squares minimization.

We formulate the fitting process as a least-squares optimization problem that minimizes the following objective function:

\begin{equation}
    J(\hat{A}, \hat{r_b}, \hat{\phi}, \hat{C}) = \int_0^{T_b} \left[ s(t) - \hat{s}(t) \right]^2 dt + \int_{T_b}^T \left[ \hat{s}(t) \right]^2 dt
    \label{eq:objective_function}
\end{equation}

Based on the above theoretical model, we can derive the analytical insights.

\subsubsection{Proposition 5 (Amplitude Attenuation)} 
When fitting a sinusoidal model to a signal containing both breathing and pause periods, the estimated amplitude $\hat{A}$ exhibits an approximately linear relationship with the breathing time ratio $T_b/T$, with oscillations due to the non‑integer‑cycle truncation.

\begin{equation}
\frac{\hat{A}}{A} \approx \frac{T_b}{T} + \delta\left(r_b, T, \phi, \frac{T_b}{T}\right)
\label{eq:amplitude_compression}
\end{equation}
where $\delta(\cdot)$ is an oscillatory correction term.

\begin{proof}
Please refer to Appendix C.1 for a detailed analysis of the amplitude attenuation.
\end{proof}

Moreover, Figs.~\ref{fig:stop1}--\ref{fig:stop2} present the simulation results of~(\ref{eq:objective_function}), with the detailed simulation setup provided in Appendix~C.1. As shown in Fig.~\ref{fig:stop1}, the fitted amplitudes oscillate around the linear attenuation trend, which is consistent with Proposition~5. Fig.~\ref{fig:stop2} further shows that the oscillation magnitude decreases as the observation window increases. To support the theoretical analysis, Appendix~C.1 presents real CSI observations of attenuation trends during breathing pauses. Based on this attenuation behavior, we establish the theoretical foundation for amplitude-threshold-based breathing pause detection in Section~IV-D.

\subsubsection{Frequency Estimation}

The frequency estimation bias can be analyzed based on the following optimization condition derived from the least-squares cost function:
\begin{align}
\frac{\partial J}{\partial \hat{r_b}} &= -4\pi \hat{A} \int_0^{T_b} \left( s(t) - \hat{s}(t) \right) t \cos(2\pi \hat{r_b} t + \hat{\phi}) \, dt \nonumber \\
&\quad + 4\pi \hat{A} \int_{T_b}^{T} \hat{s}(t) \, t \cos(2\pi \hat{r}_b t + \hat{\phi}) \, dt = 0
\label{eq:frequency_bias}
\end{align}

This optimality condition reveals two physically distinct contributions, leading to the following key insight:

\textit{Proposition 6 (Frequency Estimation): }
When fitting a sinusoidal model to a respiratory signal containing a pause interval, the breathing frequency estimate $\hat{r_b}$ is predominantly determined by the breathing segment, while the pause segment theoretically contributes no frequency-selective information and therefore introduces only limited bias.

\begin{proof}
Please refer to Appendix C.2 for a detailed analysis of the two contributions in \eqref{eq:frequency_bias}.
\end{proof}

Furthermore, Fig.~\ref{fig:stop3} presents simulation results on the effect of the observation window length, with the detailed simulation setup described in Appendix C.2. When the pause period is fixed, the average error per minute between the fitted and actual lines decreases as the total observation window length shortens. This observation motivates our use of a shorter window for breathing estimation during pause episodes in Section~IV-D.

\section{Methodology}
\begin{figure}[b]
    \centering
    \includegraphics[width=\linewidth]{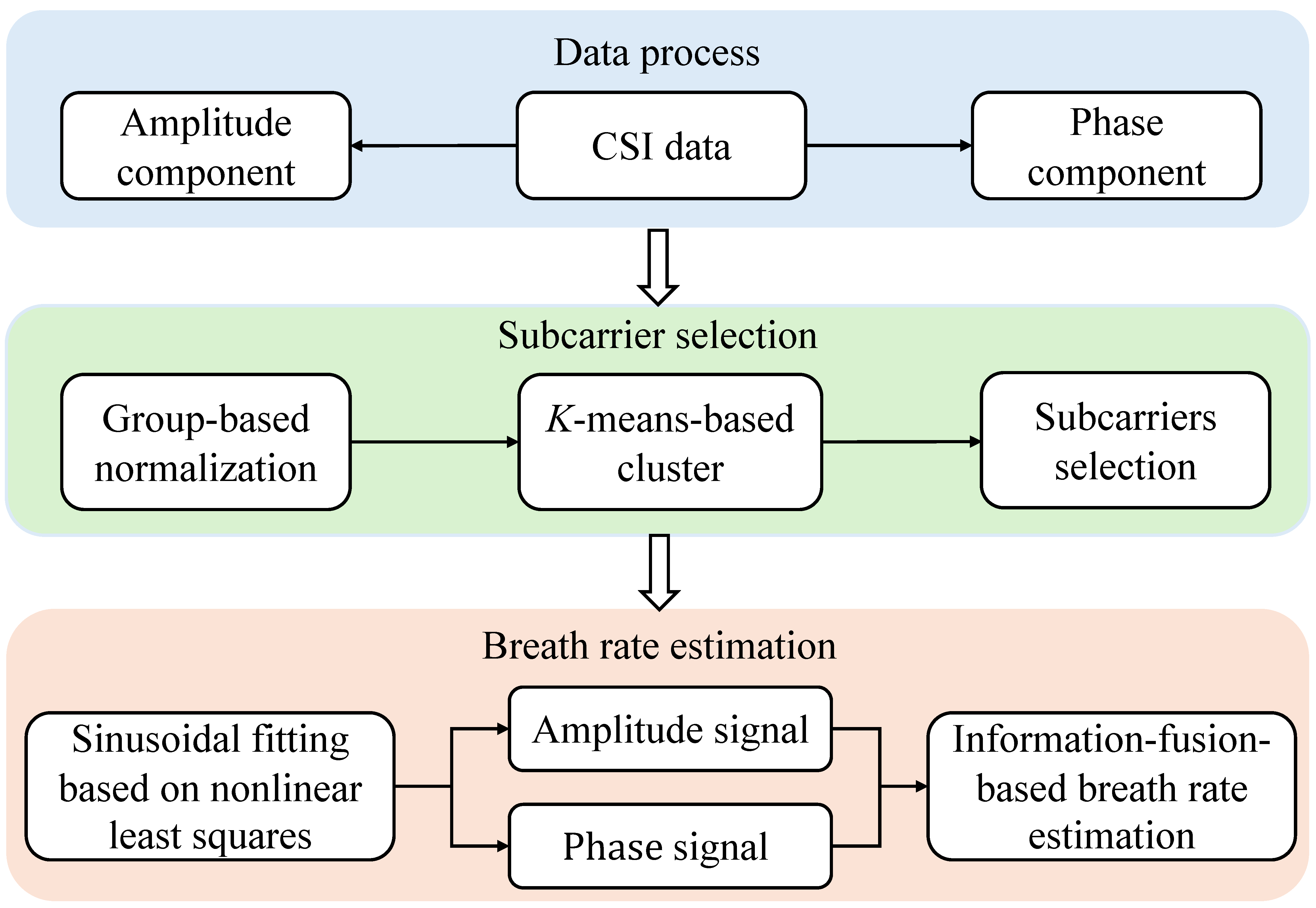}
    \caption{The flowchart of the proposed algorithm.}
    \label{fig:flowchart}
\end{figure}

Based on the theoretical foundations, this section presents the methodology of the proposed algorithm. As illustrated in Fig. \ref{fig:flowchart}, the algorithm includes three modules.

\subsection{Data Processing}
Raw CSI measurements contain both amplitude and phase information, which requires different processing steps before subsequent analysis.

\subsubsection{Amplitude Processing}
For each subcarrier $n$, the raw CSI amplitude time series $|H_n(t)|$ is extracted directly from channel measurements.

\subsubsection{Phase Processing}
The raw CSI phase is affected by packet-level phase offsets and linear phase distortions, which makes it unsuitable for direct sensing. Specifically, we follow the phase calibration procedure in \cite{long_2022_Adaptive}, which includes phase unwrapping and linear trend removal across subcarriers to reduce packet-level phase offsets and phase distortions.
\subsection{Informative Subcarrier Selection}

\subsubsection{Group-wise Normalization on Amplitudes}
Following the framework established in Section~III-B, the group-wise normalization procedure transforms $x_{m,n}$ to $y_{m,n}$:
\begin{equation}
y_{m,n} = \frac{x_{m,n} - \min_j x_{m,j}}{\max_j x_{m,j} - \min_j x_{m,j}}
\end{equation}
yielding the normalized amplitude time series $y_n(m)$ for each subcarrier. As discussed in Section~III-B, this normalization preserves the breathing periodicity of sensitive subcarriers while compressing insensitive ones to near-constant values.

\subsubsection{Four-Dimensional Feature Space}
For each subcarrier $n$, we construct a four-dimensional feature vector to enhance robustness against noise and environmental variations:

\begin{equation}
\mathbf{F}_n^{\text{(ext)}} = [\sigma_n, E_n, U_n, \sigma_{\phi,n}]^T
\end{equation}
where $U_n$ represents the autocorrelation periodicity and $\sigma_{\phi,n}$ denotes the phase standard deviation. This creates a distinctive structure in four-dimensional space that subsequent unsupervised learning-based clustering can reliably identify. The decision of this four-dimensional set is further discussed in Section V-F.

\subsubsection{Unsupervised Learning-Based Subcarrier Clustering}
Based on Section III-C, subcarriers with different sensitivity levels exhibit distinct distributions in the normalized feature space. This structural distinction enables reliable identification via unsupervised learning-based clustering.

Specifically, we apply $K$-means clustering with $K=3$ to the set of normalized feature vectors $\{\tilde{\mathbf{F}}_n\}_{n=1}^N$, where each feature is normalized to $[0,1]$ range:
\begin{equation}
\tilde{F}_{n}^{(k)} = \frac{F_n^{(k)} - \min_n F_n^{(k)}}{\max_n F_n^{(k)} - \min_n F_n^{(k)}}
\end{equation}

These clusters correspond to different levels of breathing sensitivity, i.e., high, medium, and low. To identify the set of breathing-sensitive subcarriers $\mathcal{H}^{\text{sens}}$, we select the cluster with the highest average breathing-band energy $E_n$, which corresponds to the high-sensitivity group. This energy-based criterion ensures that we retain subcarriers with the most prominent breathing signatures while excluding those dominated by noise or weak modulation.

\subsubsection{Optimal Subcarrier Selection}
From the identified sensitive subcarrier set $\mathcal{H}^{\text{sens}}$, we further select a single optimal subcarrier for subsequent breathing rate estimation. For each $n \in \mathcal{H}^{\text{sens}}$, we fit a sinusoidal model to its normalized amplitude $y_n(m)$:
\begin{equation}
\hat{y}_n(m) = \hat{A}_n \sin(2\pi \hat{r}_b t_m + \hat{\phi}_n) + \hat{C}_n
\end{equation}
where the parameters are estimated via least-squares minimization. The quality of each fit is quantified by the coefficient of determination:
\begin{equation}
R^2_n = 1 - \frac{\sum_{m=1}^M \left( y_n(m) - \hat{y}_n(m) \right)^2}{\sum_{m=1}^M \left( y_n(m) - \bar{y}_n \right)^2}
\end{equation}
The optimal subcarrier $n^*$ is selected as the one with the highest $R^2$ value:
\begin{equation}
n^* = \arg\max_{n \in \mathcal{H}^{\text{sens}}} R^2_n
\end{equation}

\begin{figure*}[ht]
    \centering
    \begin{subfigure}[b]{0.32\linewidth}
        \centering
        \includegraphics[width=0.59\linewidth]{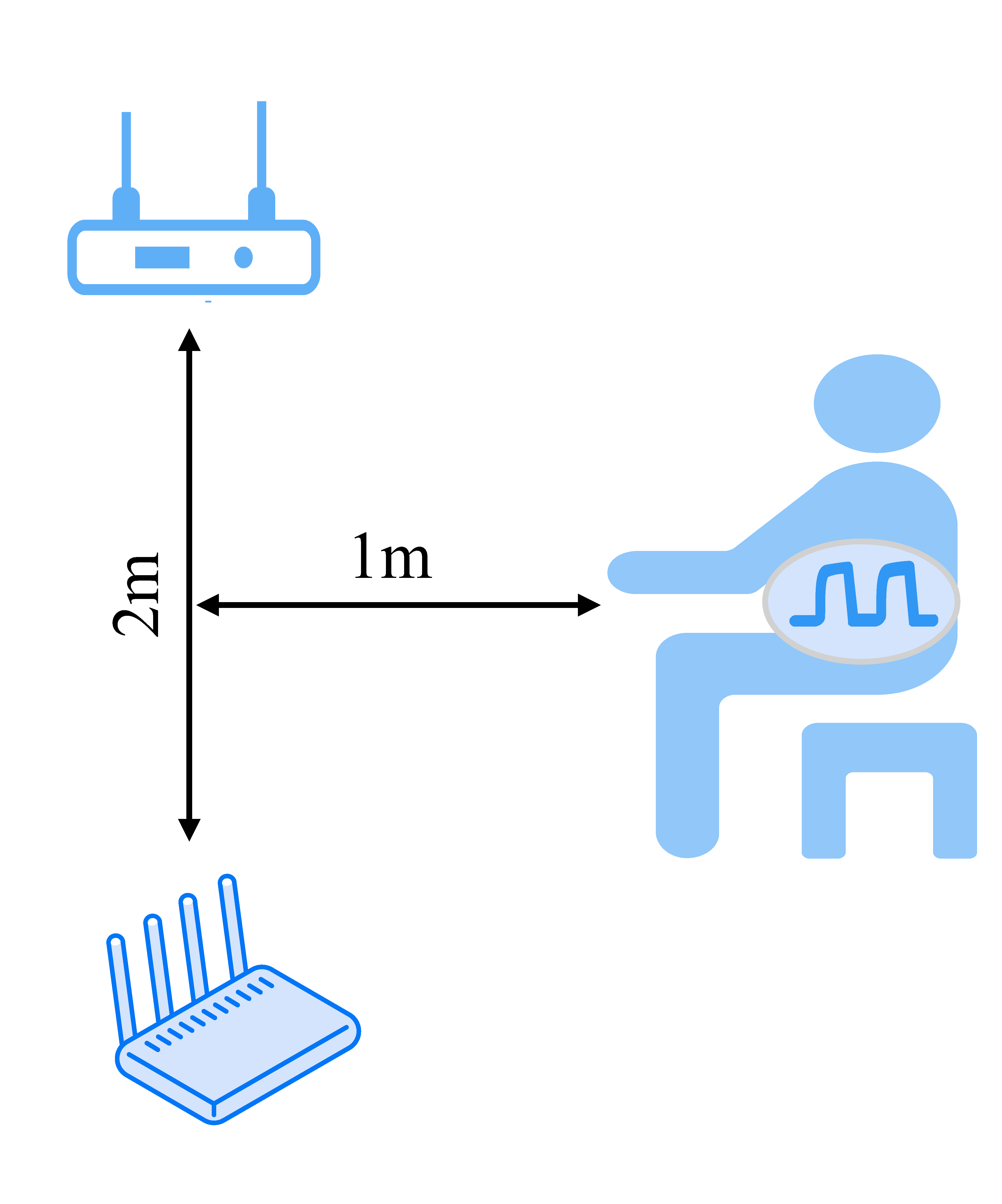}
        \caption{}
        \label{fig:scenario1}
    \end{subfigure}%
    \hfill
    \begin{subfigure}[b]{0.32\linewidth}
        \centering
        \includegraphics[width=0.59\linewidth]{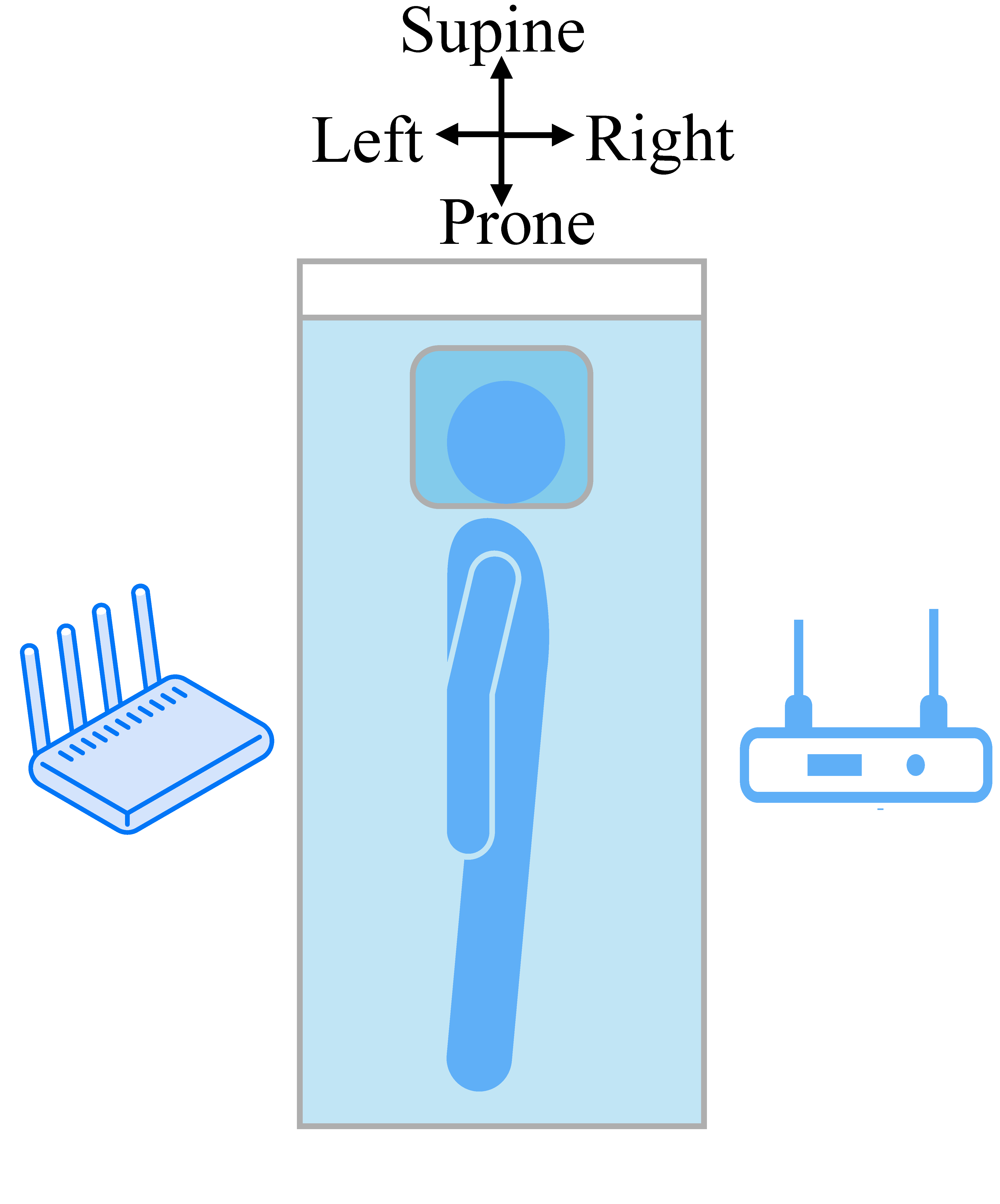}
        \caption{}
        \label{fig:scenario2}
    \end{subfigure}%
    \hfill
    \begin{subfigure}[b]{0.32\linewidth}
        \centering
        \includegraphics[width=0.59\linewidth]{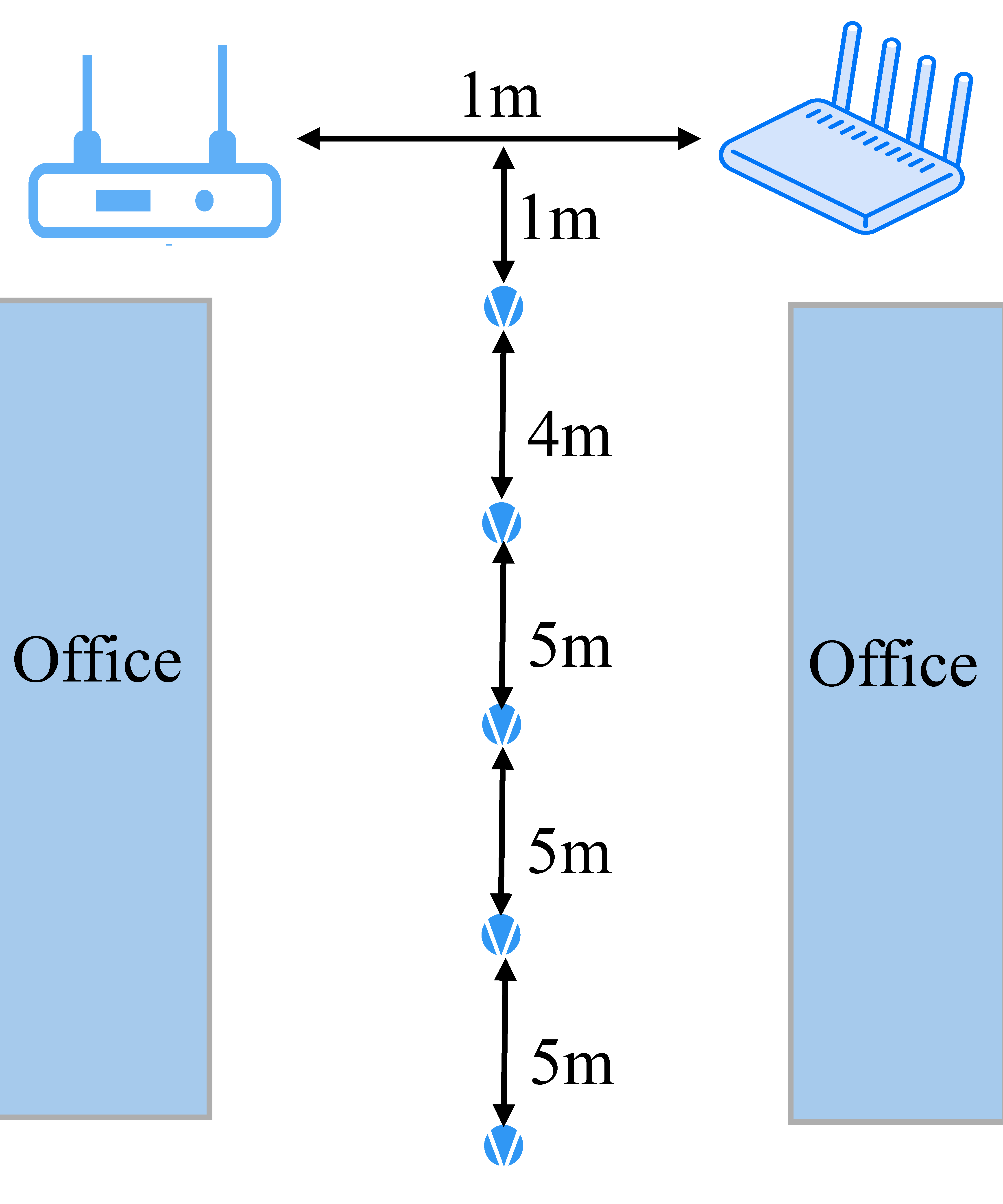}
        \caption{}
        \label{fig:scenario3}
    \end{subfigure}
    \caption{Experimental scenarios: (a) sitting person, (b) different lying postures, and (c) different distances.}
    \label{fig:scenarios}
\end{figure*}

\subsection{Breathing Rate Estimation via Information Fusion}
As discussed in Section III-D, subcarriers exhibit complementary sensitivity to breathing motions in the amplitude and phase domains. To exploit this property, we perform optimal subcarrier selection separately for each component and fuse their estimates to obtain a robust breathing rate. 

For the selected optimal subcarrier $n^*$, we perform sinusoidal fitting on both its normalized amplitude $y_{n^*}(m)$ and its calibrated phase $p_{n^*}(m)$:

\begin{align}
\hat{y}_{n^*}(m) &= \hat{A}_{\text{amp}} \sin(2\pi \hat{r}_{\text{amp}} t_m + \hat{\phi}_{\text{amp}}) + \hat{C}_{\text{amp}} \\
\hat{p}_{n^*}(m) &= \hat{A}_{\text{phase}} \sin(2\pi \hat{r}_{\text{phase}} t_m + \hat{\phi}_{\text{phase}}) + \hat{C}_{\text{phase}}
\end{align}
where $\hat{r}_{\text{amp}}$ and $\hat{r}_{\text{phase}}$ are the breathing frequencies estimated from amplitude and phase, respectively.

For each fit, we compute the corresponding $R^2$ values: $R^2_{\text{amp}}$ and $R^2_{\text{phase}}$. The final breathing rate estimate $\hat{r}_b$ is obtained by selecting the candidate with the higher fit quality:
\begin{equation}
\hat{r}_b = 
\begin{cases}
\hat{r}_{\text{amp}}, & \text{if } R^2_{\text{amp}} \geq R^2_{\text{phase}}, \\
\hat{r}_{\text{phase}}, & \text{otherwise}.
\end{cases}
\end{equation}

This fusion strategy ensures that the most reliable modality for the selected subcarrier contributes to the final estimate. 

We evaluate performance using accuracy and mean absolute error (MAE) \cite{ge_2025_WiRe,zeng_2019_Farsense}. These metrics are averaged over all test samples $M$ to obtain the final performance indicators:

\begin{equation}
\text{Accuracy} = \left(1 - \frac{|\hat{r}_b - r_b|}{r_b}\right) \times 100\%
\end{equation}

\begin{equation}
\text{MAE} = \frac{1}{M} \sum_{i=1} ^{M} |\hat{r}_b - r_b|
\end{equation}
where $\hat{r}_b$ and $r_b$ are the estimated and ground truth breathing rates, respectively.

\subsection{Detection Method of Breathing Pauses}

Based on the theoretical analysis in Section III-E, we further develop a threshold-based pause detection mechanism. The derived amplitude attenuation relationship provides the theoretical basis for detecting pauses via amplitude thresholding, while the frequency estimation analysis ensures reliable breathing rate estimation even during pause episodes. Furthermore, simulation results in Section III-E show that a shorter observation window reduces estimation error, which motivates the adoption of a 3-second sliding window for breath estimation during pause detection.

Specifically, the detection process operates as follows. For each window, we fit a sinusoidal model to the selected breathing-sensitive subcarriers and extract the estimated amplitude $\hat{A}$. The reference amplitude $A_{\text{ref}}$ is obtained as the mean of $\hat{A}$ of normal breathing (free of pauses) from a normal-breathing calibration period. For subsequent testing, a breathing pause is declared when the estimated amplitude falls below a fixed proportion of this reference, i.e., $\hat{A} < \eta \cdot A_{\text{ref}}$, where $\eta = 0.3$ is empirically determined. The sensitivity analysis of $\eta$ is conducted in Section~V-C. Beyond detection, we estimate pause duration based on the linear relationship in Proposition~5:

\begin{equation}
T_a = T - T_b = T\left(1 - \frac{\hat{A}}{A_{ref}}\right)
\end{equation}
where $A_{ref}$ is the reference amplitude during normal breathing periods.

To evaluate detection performance, we define the false-detection rate (FDR) and missed-detection rate (MDR) using event-level measures of false alarms and missed pauses. FDR measures false alarms, while MDR measures missed detections, which are calculated as follows:
\begin{equation}
\text{FDR} = \frac{\text{FD}}{\text{Detected Pauses}} \times 100\%
\end{equation}

\begin{equation}
\text{MDR} = \frac{\text{MD}}{\text{Actual Pauses}} \times 100\%
\end{equation}
where FD represents the number of false detections, MD represents the number of missed detections, Detected Pauses is the total number of pause events detected by the algorithm, and Actual Pauses is the total number of true pause events in the ground truth.

For duration estimation accuracy, we compute the MAE between the estimated and ground-truth durations:

\begin{equation}
\text{Duration Error} = \frac{1}{P}\sum_{i=1}^{P} |\hat{T}_a^{(i)} - T_a^{(i)}|
\end{equation}
where $\hat{T}_a^{(i)}$ and $T_a^{(i)}$ are the estimated and actual durations for the $i$-th pause event, respectively, and $P$ is the total number of detected events.

\subsection{Computational Complexity and Real-Time Performance}
The computational complexity of the proposed algorithm is dominated by three steps: group-wise normalization (\(O(N \cdot G)\)), feature extraction (\(O(N \cdot G)\)), and $K$-means clustering (\(O(N \cdot K \cdot G)\)), where \(N\) is the number of subcarriers, \(K\) is the number of clusters, and \(G = M / L\) is the number of time groups after averaging, where \(M\) is the total number of received packets, \(L\) is the averaging length per group. Thus, the overall complexity is \(O(N \cdot K \cdot G)\), linear in the input size.

To evaluate practical, real-time performance, we measured execution time on a PC with an Intel i7-13700KF CPU and 16 GB RAM using MATLAB. The average processing time for the algorithm to process one minute of CSI data is 3.26 seconds, and the average packet processing time is approximately 1 ms. Additionally, the memory footprint remained below 100 MB. These results suggest that the proposed method is computationally feasible for continuous respiratory monitoring and practical deployment in home healthcare environments.

\subsection{Ethics Approval of Data Collection}
This study involving human participants was reviewed and approved by the Institutional Review Board of The Hong Kong Polytechnic University (Approval No. HSEARS20250506004).

\begin{figure*}[ht]  
    \centering
    \begin{minipage}[b]{0.32\linewidth} 
        \centering
        \includegraphics[width=\linewidth]{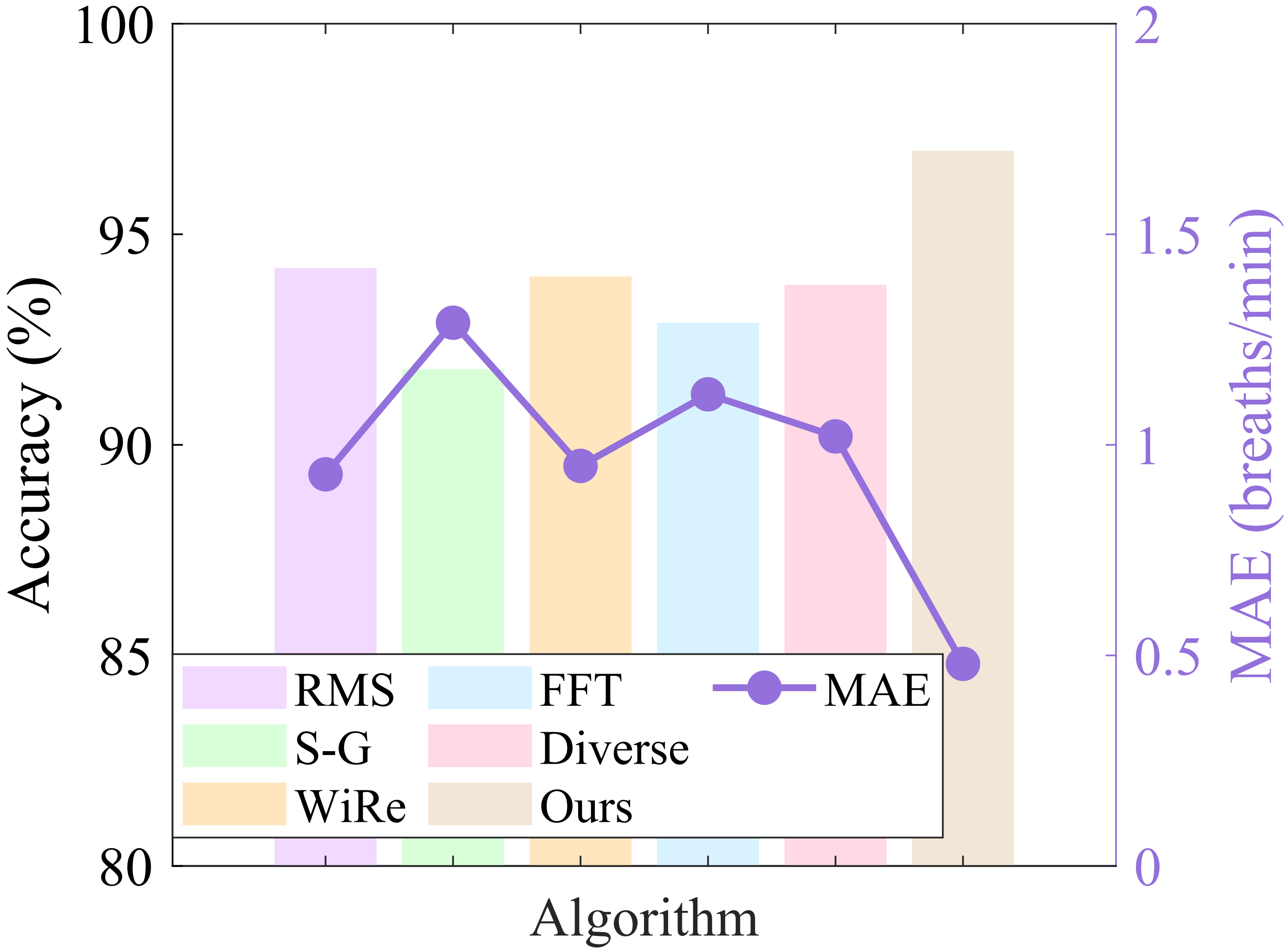} 
        \caption{Algorithm performance comparison under normal breathing.}
        \label{fig:result1}
    \end{minipage}%
    \hfill
    \begin{minipage}[b]{0.32\linewidth} 
        \centering
        \includegraphics[width=\linewidth]{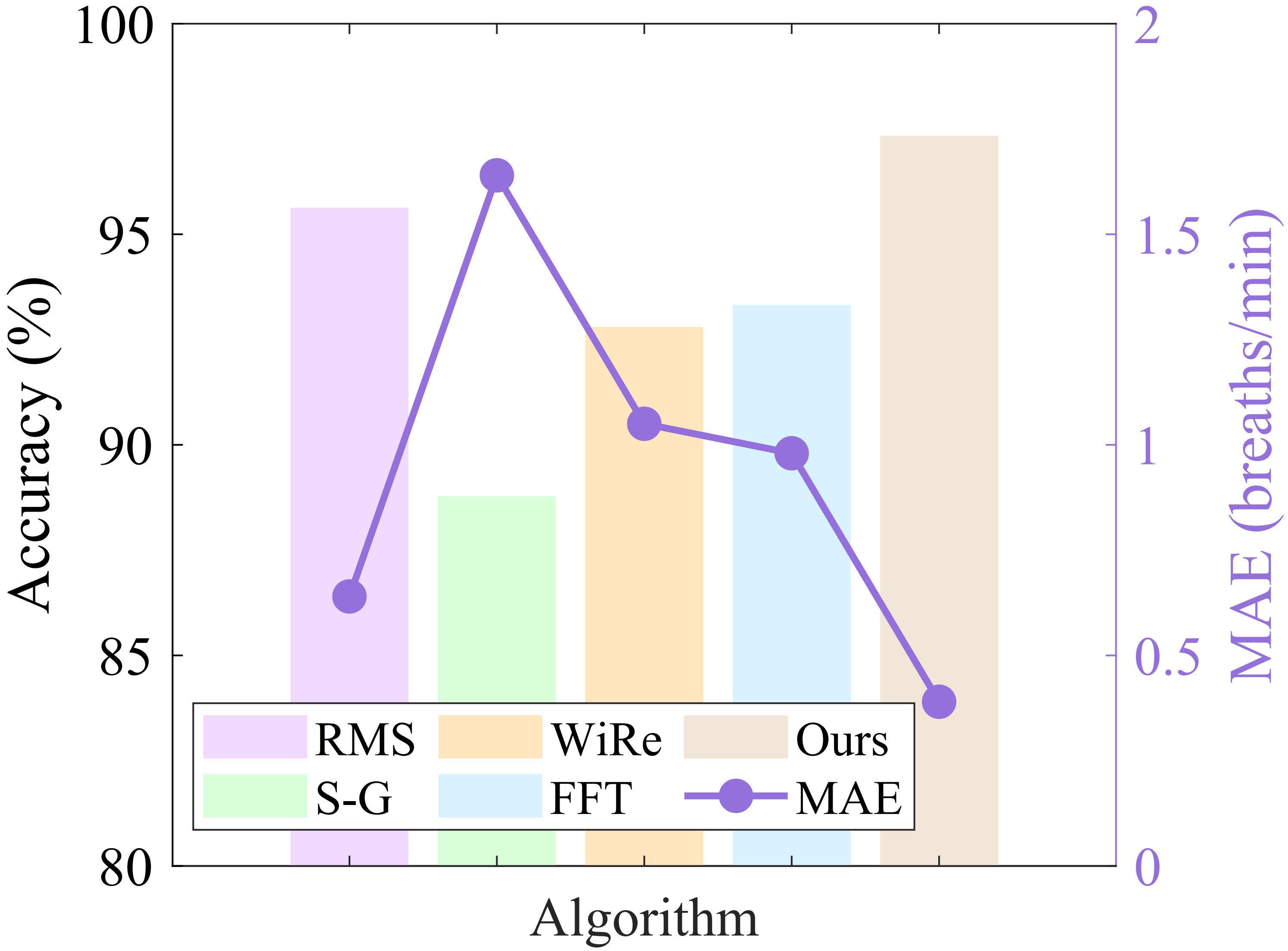}  
        \caption{Algorithm performance comparison on the public dataset.}
        \label{fig:result2}
    \end{minipage}%
    \hfill
    \begin{minipage}[b]{0.32\linewidth} 
        \centering
        \includegraphics[width=0.9\linewidth]{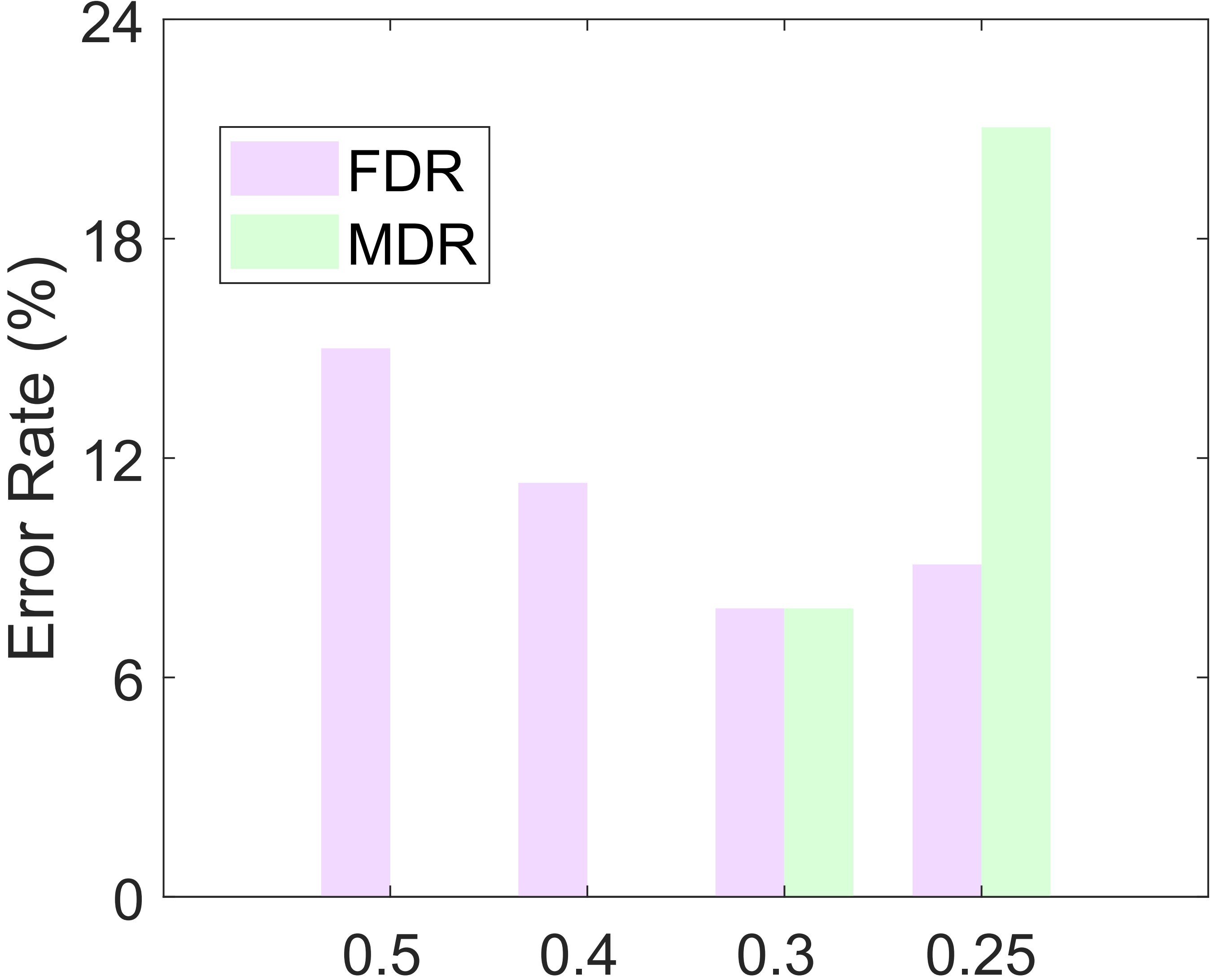}  
        \caption{Performance of pause detection under different proportion thresholds $\eta$.}
        \label{fig:result8}
    \end{minipage}%

\end{figure*}

\section{EXPERIMENTAL RESULTS AND DISCUSSION}
This section evaluates the algorithm's performance across various scenarios.

\subsection{Experiment Setup}

The experimental setup is illustrated in Fig.~\ref{fig:scenarios}, encompassing indoor environments, various lying postures and corridors. Our data acquisition system consists of a Raspberry Pi 4B running the Nexmon framework, which extracts 256 CSI subcarriers from 5 GHz Wi-Fi devices operating at 80 MHz bandwidth. Following \cite{ge_2025_WiRe}, we use the beacon signal to collect CSI data at 50 Hz. More details of the data collection system can be found in our previous work \cite{tang_2024_bicsi}. Ground truth respiratory data were recorded using the Neulog Respiration Monitor Belt Logger Sensor. All collected data were processed using MATLAB 2023a.

During data collection, four volunteers participated in 58 tests (28 with normal breathing and 30 with breath-holding events) to evaluate algorithm performance. To further validate generality, we also evaluated our algorithms on a public dataset comprising 30 tests from \cite{Alzaabi_2024_eldercare}. For respiratory pause assessment, participants performed normal breathing interspersed with voluntary breath-holding events. These episodes naturally varied in duration, ranging from approximately 2 to 10 seconds, reflecting the variability of unscripted behavior.

\subsection{Evaluation of Overall Algorithm Performance}

To evaluate the effectiveness of the proposed algorithm, we test it across diverse environments and compare its performance with several related algorithms. The Savitzky-Golay (S-G) filter and Fast Fourier Transform (FFT) represent traditional approaches that directly estimate breathing rate from raw subcarrier signals \cite{liu_2021_FFT}. Diverse employs a combination of amplitude and phase information to achieve robust estimation \cite{li_2022_Diversense}. RMS \cite{Ge_2024_RMS} and WiRe \cite{ge_2025_WiRe} are subcarrier-aware approaches that adopt breathing energy and variance as subcarrier selection metrics, respectively.

First, we compare the performance of all methods in estimating breathing rate under normal breathing, as shown. As shown in Fig.~\ref{fig:result1}, traditional methods (FFT and S-G filter) achieve accuracies below 92\% and MAEs above 1 bpm. In contrast, subcarrier-aware methods (ours, WiRe, and RMS) consistently achieve average accuracies above 94\% and MAEs below 1 bpm, which demonstrates the effectiveness of leveraging subcarrier heterogeneity. Among these subcarrier-aware approaches, our method achieves the highest average accuracy of 97\% and the lowest MAE of 0.48 bpm, which outperforms the second-best algorithm by approximately 2.8\% in accuracy and 0.45 bpm in MAE. This improvement demonstrates that our framework outperforms single-metric subcarrier selection methods such as WiRe and RMS.

As shown in Fig~\ref{fig:result2}, we further evaluate these algorithms on a public dataset. Diverse is excluded from this comparison because it requires phase alignment across subcarriers, while the public dataset provides only amplitude information. Fig~\ref{fig:result2} shows that our algorithm maintains superior performance with 97\% accuracy and 0.4 bpm MAE. Through subcarrier selection, RMS achieves the second-best performance with 95\% accuracy and 0.6 bpm MAE, followed by WiRE and FFT with accuracies around 92\% accuracy and MAEs of around 1 bpm. The S-G filter exhibits the lowest performance among all methods, which achieves only 88\% accuracy with 1.6 bpm MAE. Overall, the consistent performance across both our collected data and the public dataset suggests the robustness and generalization capability of the proposed algorithm.

\begin{figure*}[ht]  
    \centering
    \begin{minipage}[b]{0.32\linewidth}  
        \centering
        \includegraphics[width=0.9\linewidth]{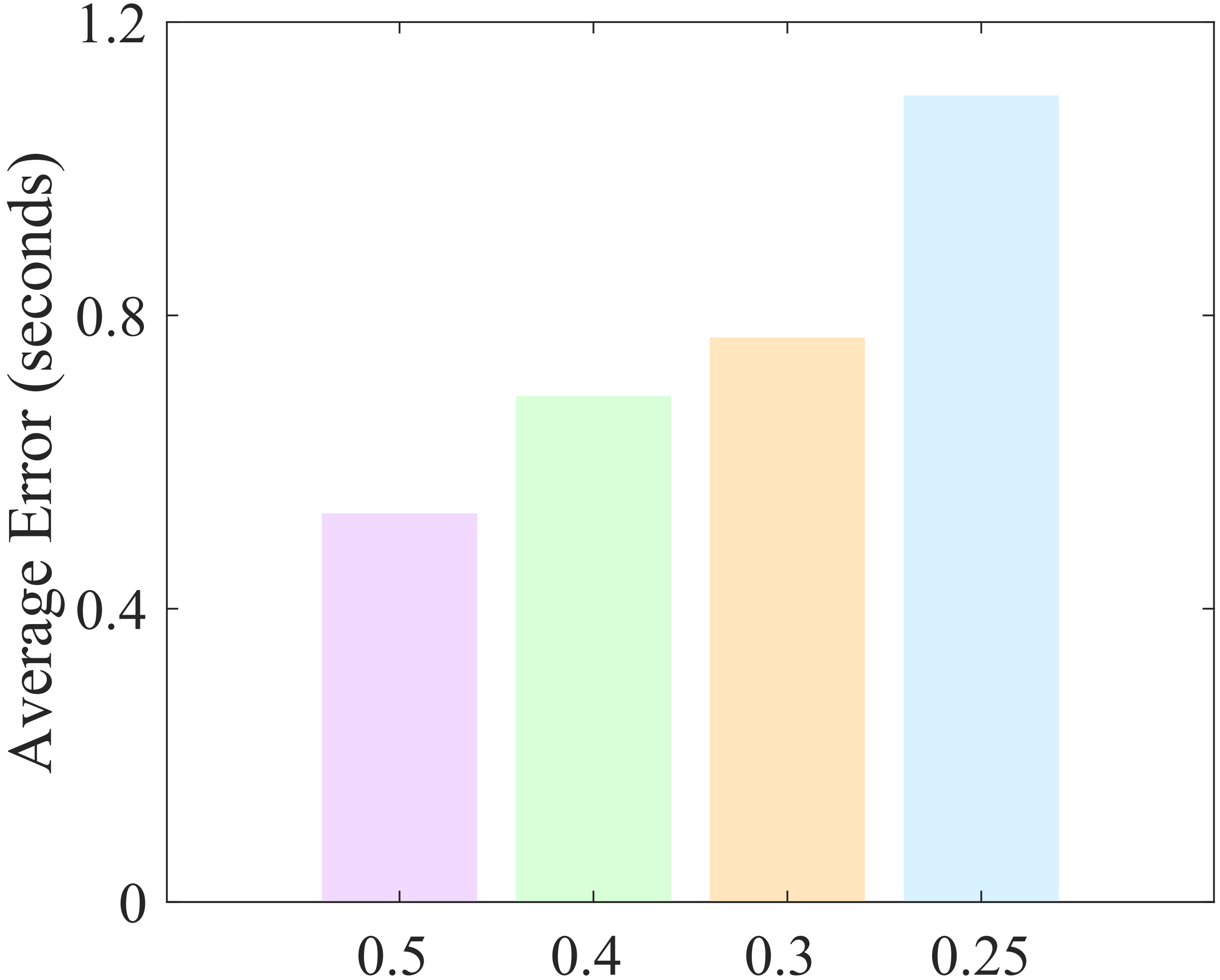} 
        \caption{Average time estimation error per detected pause.}
        \label{fig:result9}
    \end{minipage}%
        \hfill
    \begin{minipage}[b]{0.32\linewidth} 
        \centering
        \includegraphics[width=\linewidth]{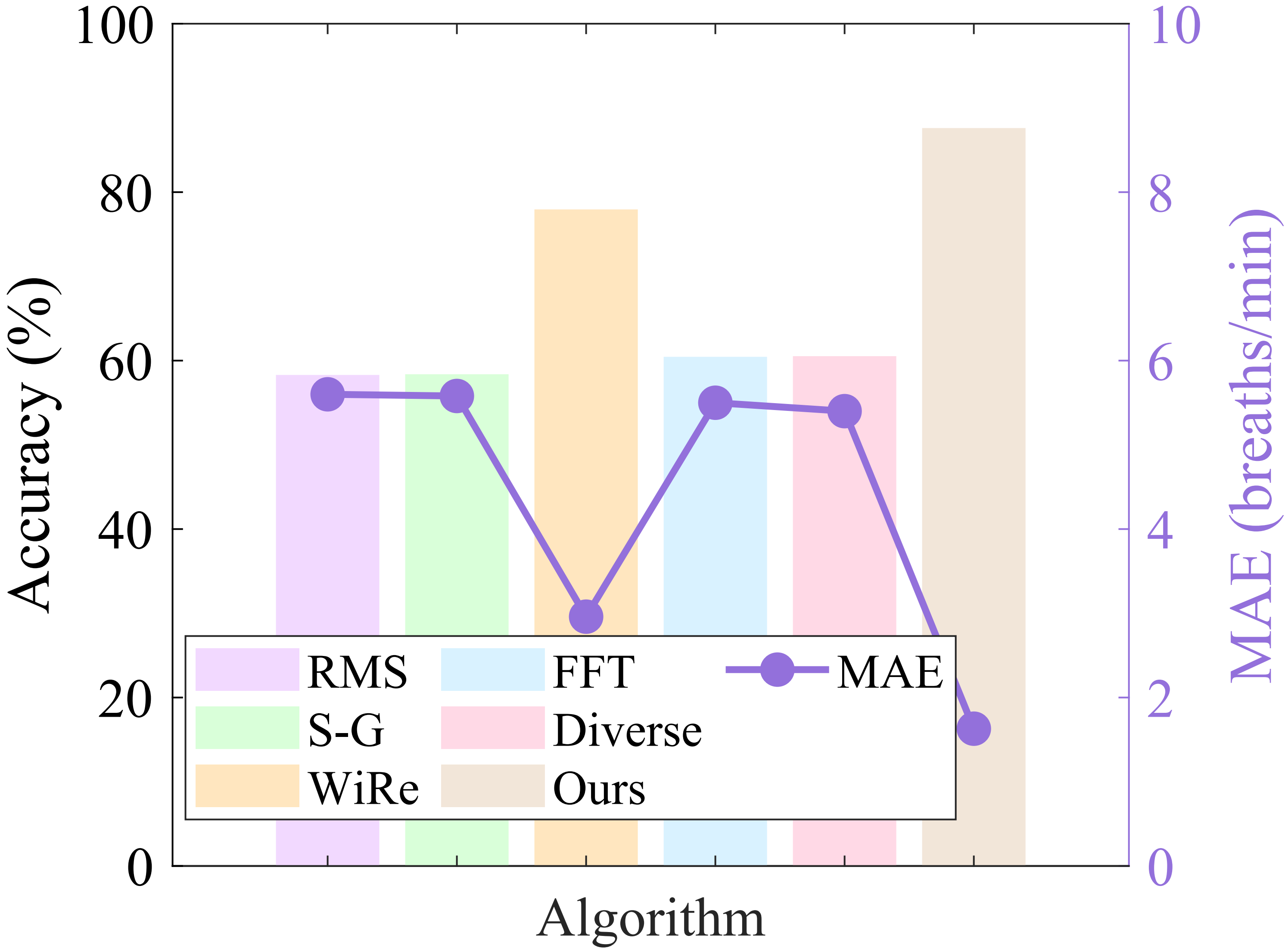} 
        \caption{Algorithm performance under breathing pause conditions.}
        \label{fig:result7}
    \end{minipage}%
        \hfill   
     \begin{minipage}[b]{0.32\linewidth}  
        \centering
        \includegraphics[width=\linewidth]{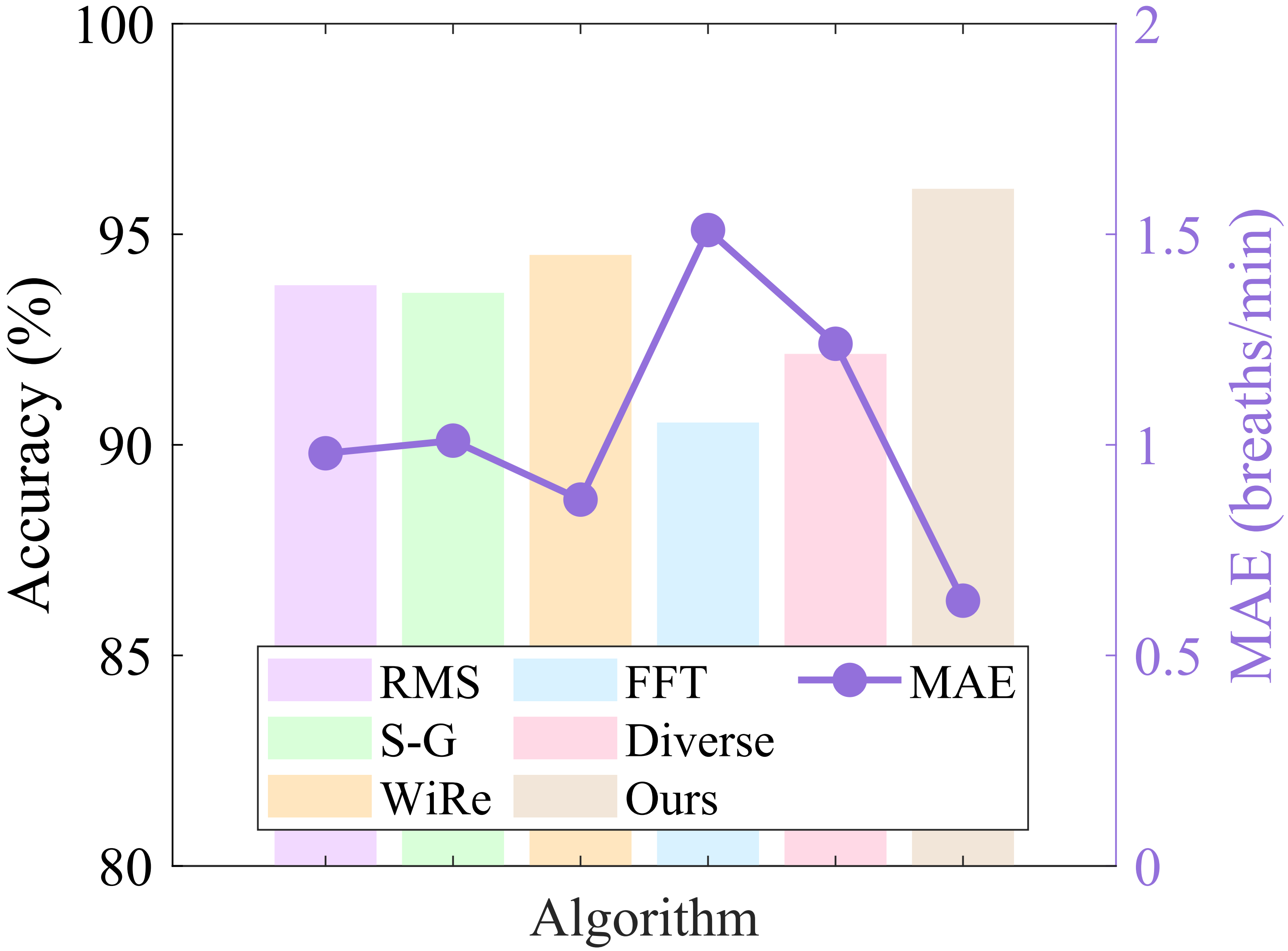} 
        \caption{Algorithm performance comparison across lying postures.}
        \label{fig:result3}
    \end{minipage}

\end{figure*}

\subsection{Evaluation Under Breathing Pauses}
Based on Section~IV-D, this section evaluates the proposed pause detection method using amplitude thresholding. Fig.~\ref{fig:result8} presents the detection performance under different values of the proportion threshold $\eta$, where the detection threshold is defined as $\eta \cdot A_{\text{ref}}$. Lowering $\eta$ from $0.5$ to $0.3$ reduces FDR from $15\%$ to $8\%$ while maintaining MDR at approximately $8\%$. However, further decreasing $\eta$ to $0.25$ causes FDR to slightly rise to $9\%$ and MDR to increase sharply to $22\%$. A favorable trade-off is achieved at $\eta = 0.3$, where both FDR and MDR remain below $10\%$. Therefore, $\eta = 0.3$ is selected as the empirical threshold in our experiments because it provides a favorable balance between false detections and missed detections while maintaining both metrics below $10\%$. Threshold recalibration may be required for different deployment scenarios.

Building on the detection results, we further evaluate pause duration estimation to validate the theoretical analysis in Section~III-E.  As discussed, the estimated amplitude exhibits an approximately linear relationship with the breathing time ratio. Fig.~\ref{fig:result9} shows the average duration estimation error under different values of $\eta$. The results show that errors increase progressively as the threshold decreases. With the reference amplitude set to $\eta = 0.5$, the estimation error is around 0.5~s, which corresponds to approximately 18\% of the 3-second observation window used for pause analysis. This error is higher than the theoretical simulation results in Section~III-E because voluntary breath-holds often contain residual chest movements that introduce amplitude fluctuations. These findings demonstrate the potential of amplitude-based duration estimation for clinically relevant applications, such as apnea detection.

Beyond detection, we evaluate breathing rate estimation during detected pause episodes. As shown in Fig.~\ref{fig:result7}, the estimation accuracy of all algorithms declines under pause conditions, which indicates the negative impact of breathing pauses. Among these, our proposed algorithm achieves the best performance, with an MAE of 1.6 bpm and an accuracy of 88\%, which significantly outperforms all baseline methods by over 9\%. WiRe ranks second with an accuracy of 78\% and an MAE of 3 bpm, while other methods exhibit substantially higher errors, with MAEs exceeding 5 bpm. These results suggest the effectiveness of our pause-specific estimation strategy.

\subsection{Impact of Different Postures}

Respiration monitoring during lying states is common in daily life, while different postures can affect signal quality and estimation accuracy \cite{ge_2025_WiRe,liu_respiration}. To further evaluate algorithm performance under such conditions, we conduct experiments in four lying postures: right-side, left-side, supine, and prone, as illustrated in Fig.~\ref{fig:scenario2}.

Fig.~\ref{fig:result3} illustrates the average performance for each algorithm across these postures. As shown in the figure, our method achieves the highest average accuracy of 96\% and the lowest average MAE of 0.63 bpm across posture variations, which outperforms the second-best algorithm (WiRe) by 1.6\% in accuracy and 0.24 bpm in MAE. Among other subcarrier-selective approaches, WiRe and RMS achieve accuracies of around 93\% with MAEs exceeding 0.9 bpm, which outperforms the remaining baselines. By contrast, FFT exhibits the lowest average performance (around 90\%), due to its large fluctuations across postures. These results highlight the robustness of subcarrier-selective approaches for lying scenarios, with our method consistently outperforming its counterparts.

\subsection{Impact of Testing Distances}

\begin{figure*}[ht]  
    \centering
        \begin{minipage}[b]{0.32\linewidth}  
        \centering
        \includegraphics[width=\linewidth]{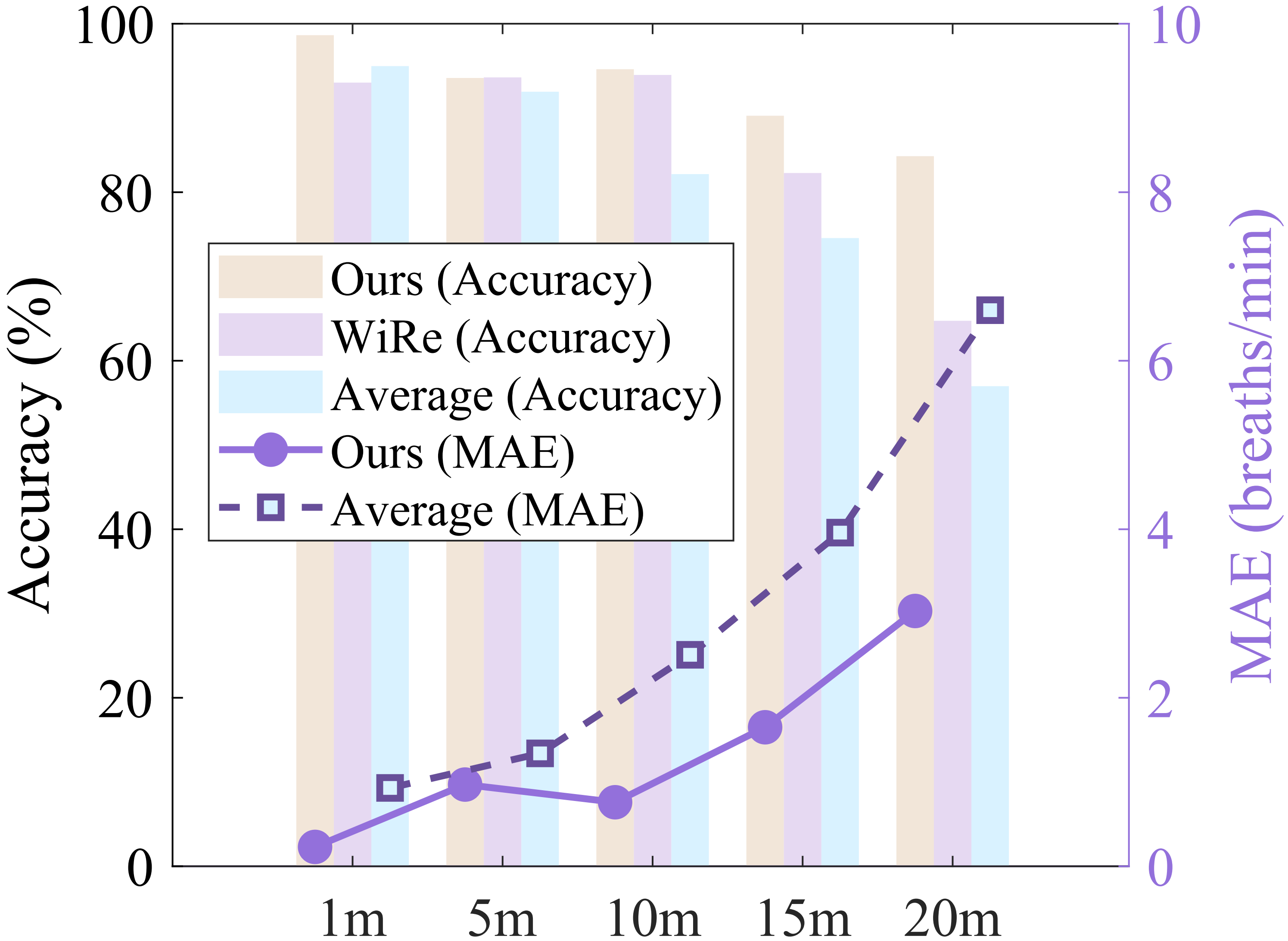} 
        \caption{Algorithm performance at different distances.}
        \label{fig:result6}
    \end{minipage}
    \hfill
     \begin{minipage}[b]{0.32\linewidth} 
        \centering
        \includegraphics[width=\linewidth]{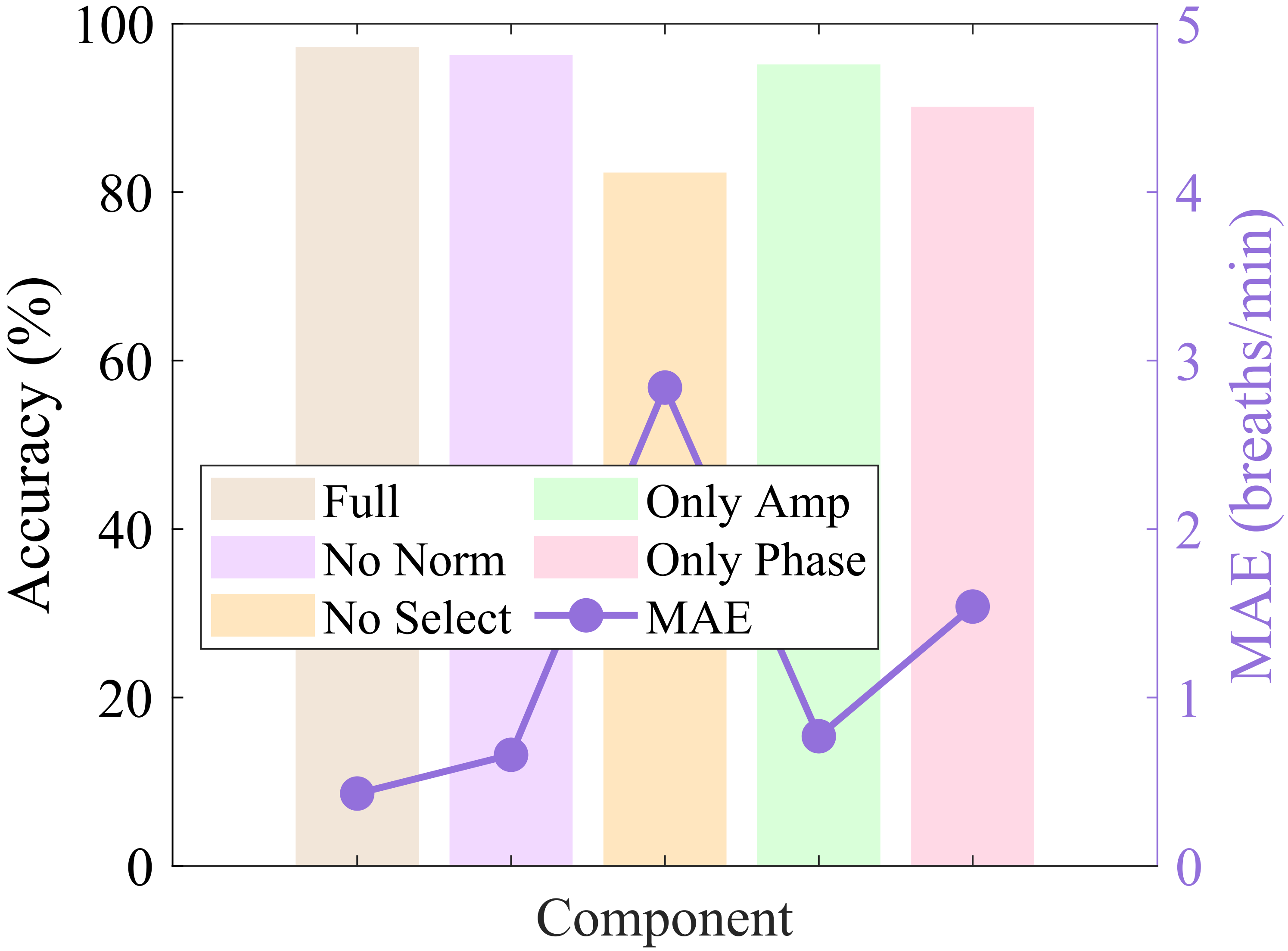} 
        \caption{Ablation study of different framework components.}
        \label{fig:result4}
    \end{minipage}%
    \hfill
    \begin{minipage}[b]{0.32\linewidth} 
        \centering
        \includegraphics[width=\linewidth]{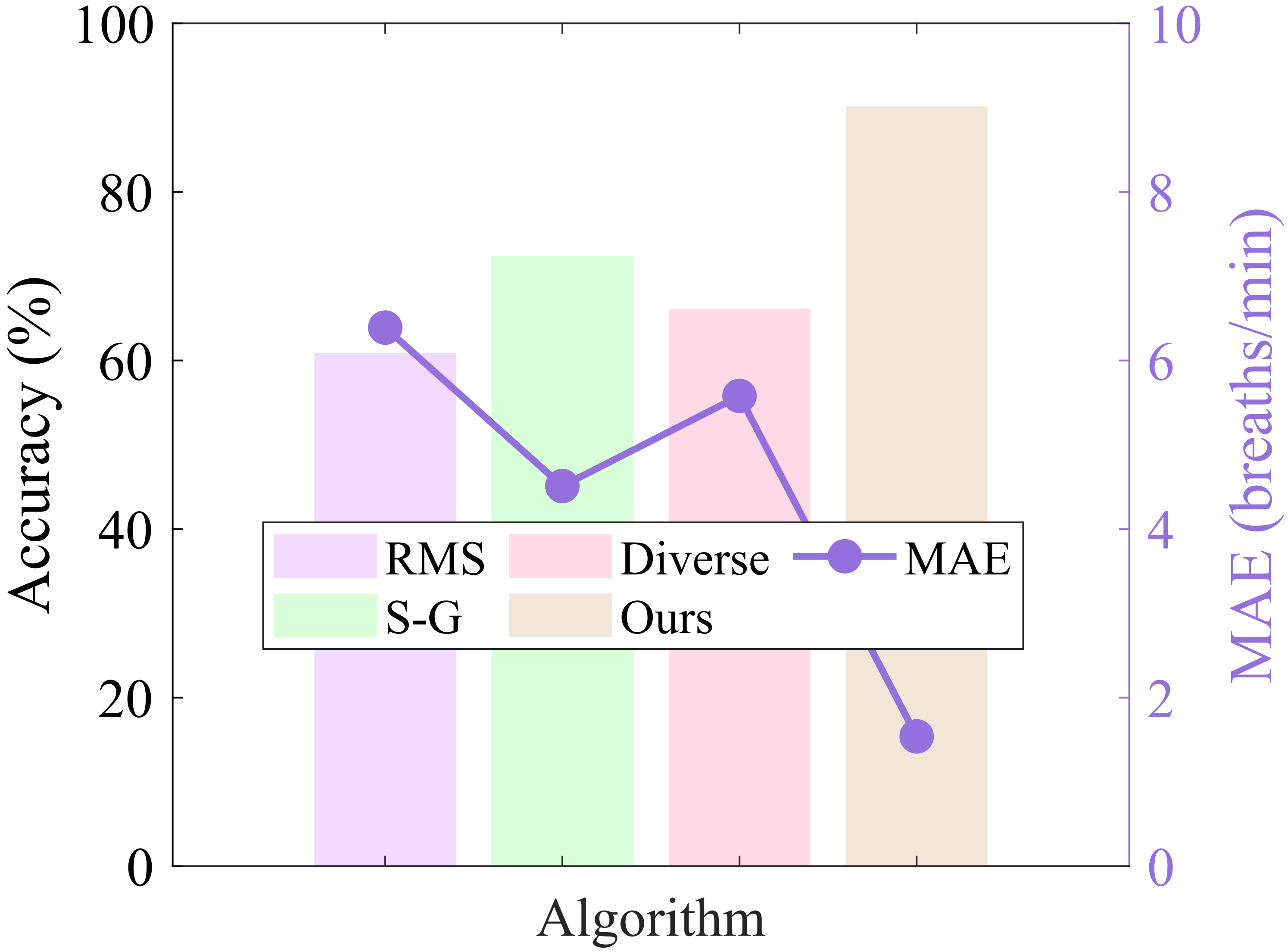}  
        \caption{Average correlation values among the feature metrics.}
        \label{fig:result5}
    \end{minipage}%
\end{figure*}

To evaluate the robustness under weak respiratory modulation conditions caused by signal attenuation, we conduct experiments in a corridor environment at distances ranging from 1 m to 20 m. The details of the experimental setup are illustrated in Fig.~\ref{fig:scenario3}. Fig.~\ref{fig:result6} presents the MAE and accuracy results for our proposed algorithm alongside the average performance of all methods.

The experimental results reveal a clear distance-dependent degradation in breathing estimation performance. To isolate the effect of distance from algorithm-specific variations, we examine the average performance across all algorithms as a baseline. The average accuracy remains above 90\% within 5 m, drops to around 80\% at 10 m, and falls below 60\% at 20 m, which confirms the substantial impact of signal attenuation at longer distances. Against this baseline, our proposed algorithm exhibits greater resilience, maintaining accuracy above 90\% within 10 m and 84\% at 20 m while consistently outperforming the average across all distances. As shown in Fig.~\ref{fig:result6}, our method also surpasses the second-best algorithm WiRe, particularly at 20 m, where it achieves 84\% accuracy compared to 65\%. These results suggest that the proposed framework remains effective under distance-induced signal degradation and weak respiratory modulation conditions.

\subsection{Impact Analysis of Key Components}

\subsubsection{Ablation tests} 
We conduct ablation experiments to evaluate the individual contributions of the key components of our framework. As shown in Fig. \ref{fig:result4}, removing any module degrades performance, demonstrating the contribution of each component to the overall framework. Specifically, removing normalization results in a 1.3\% drop in accuracy. Replacing sinusoidal fitting with FFT reduces accuracy by 2.4\%. More importantly, using one arbitrarily selected subcarrier (without adaptive selection) degrades accuracy by 15\%, which highlights the importance of the proposed subcarrier-selection mechanism. Finally, compared with amplitude-only and phase-only approaches, the proposed fusion method improves accuracy by 2\% and 7\%, respectively, which indicates the complementary benefits of information fusion.

\subsubsection{Phase-Only Performance Advantage}
Our method also demonstrates a significant advantage in phase-only estimation. As shown in Fig.~\ref{fig:result5}, even with phase information alone, our method achieves 90\% accuracy with an MAE of 1.5 bpm, outperforming other phase-only approaches by over 17\% in accuracy and more than 2 bpm in MAE. (WiRe and FFT are excluded as they are designed for amplitude-based estimation.) This advantage further suggests the effectiveness of our subcarrier selection and processing strategy.

\subsubsection{Parameter and Setting Analysis}
The algorithm's parameter settings are analyzed as follows.

\textit{1. Grouping window size.} To evaluate parameter sensitivity, we tested averaging windows of 0.4, 1, 2, and 4 seconds. The corresponding breathing-rate estimation accuracies were 94\%, 97\%, 89\%, and 89\%, respectively. Among the tested settings, the one-second window achieved the highest accuracy and was therefore used throughout this work.

\textit{2. Feature matrix.} To verify the effectiveness of the four selected features, we remove each feature individually from the four-dimensional set \([\sigma_n, E, U_n, \sigma_{\phi,n}]\) and re-evaluate breathing rate estimation. As shown in Table~\ref{tab:feat_ablation}, the MAE increases by over 0.13 bpm in each case, thereby confirming that all four features contribute meaningfully to performance.

\begin{table}[ht]
\centering
\caption{The MAE increase in ablation studies of the feature matrix.}
\label{tab:feat_ablation}
\begin{tabular}{cc}
\toprule
Removed feature & MAE increase (bpm) \\
\midrule
\(\sigma_n\) (STD) & 0.13 \\
\(E_n\) (breathing-band energy) & 0.20 \\
\(U_n\) (autocorrelation) & 0.24 \\
\(\sigma_{\phi,n}\) (phase STD) & 0.21 \\
\bottomrule
\end{tabular}
\end{table}

\textit{3. Clustering number \(K\).} The proposed unsupervised clustering method uses \(K=3\) to group subcarriers by sensitivity. To justify this choice, we evaluate both clustering quality and sensing performance under \(K=2\) and \(K=4\). As shown in Table~\ref{tab:K_impact}, \(K=3\) achieves the highest silhouette coefficient (0.67), which indicates the best cluster separation. Although \(K=2\) yields comparable estimation accuracy, its silhouette coefficient is lower than that of \(K=3\). Increasing $K$ to 4 reduces both the silhouette coefficient and estimation accuracy. Therefore, \(K=3\) is selected as the optimal configuration.

\begin{table}[ht]
\centering
\caption{Breathing estimation accuracy and silhouette coefficients for different \(K\) values.}
\label{tab:K_impact}
\begin{tabular}{cccc}
\toprule
\(K\) Value & 2 & 3 & 4 \\
\midrule
Accuracy  & 97\% & 97\% & 96\% \\
Silhouette & 0.65 & 0.67 & 0.60 \\
\bottomrule
\end{tabular}
\end{table}

\textit{4. Threshold cross-validation.} To further validate the generalizability of the threshold \(\eta = 0.3\) for pause detection, we conducted an additional ten tests in a new indoor environment with two new volunteers (transceiver distance: 1.5 m; participant seated midway). Using the same \(\eta = 0.3\) without the recalibration, the FDR and MDR remain below 10\% (7\% and 4\%, respectively). These results suggest that the chosen threshold generalizes reasonably well across different environments and subjects. If possible, threshold recalibration is still recommended for different deployment scenarios.

\subsection{Limitations and Future Work}

Despite the promising results, several limitations should be discussed. First, the current framework assumes a single stationary breathing source and relies on empirically determined thresholds. Additionally, the theoretical analysis relies on simplifying assumptions, such as the dominance of target motion over noise, which may not fully reflect the complexity of real-world environments (e.g., multipath fading, motion interference). Compared to the stationary environments studied in this work, these complex scenarios may require more refined models.

Furthermore, the theoretical pause analysis and sinusoidal fitting framework assume abrupt transitions between breathing and silence and a single dominant breathing frequency. These assumptions may not hold under more complex conditions, such as irregular respiration, apnea recovery, or motion disturbances. Consequently, the pause-detection mechanism may require further refinement to handle such conditions. Moreover, because the empirical threshold is derived from our current dataset and may vary across deployment environments, recalibration is recommended for practical deployment.

Despite these limitations, the theoretical insights and experimental results in this paper provide a foundation for further research in contact-free respiratory monitoring. Future work will extend the framework to multi-person sensing, investigate adaptive thresholding strategies, explore time-frequency approaches to handle nonstationary breathing patterns, and further study the physical mechanisms underlying subcarrier heterogeneity. Additionally, we plan to validate the method on more diverse real-world data and develop online calibration procedures for practical deployment in home healthcare scenarios.

\section{Conclusion}

This paper extends traditional Wi-Fi-based breathing sensing by accounting for subcarrier-dependent heterogeneity in CSI measurements. Unlike conventional approaches that treat all subcarriers uniformly, our theoretical analysis establishes a subcarrier-dependent model that captures heterogeneous respiratory sensitivities across subcarriers. Based on this model, we propose an unsupervised clustering method to identify informative subcarriers and a sinusoidal-fitting scheme for respiratory-rate estimation. Experimental results show that the proposed method achieves 97\% accuracy and reduces the MAE by 0.45 bpm compared with baseline methods. The framework also demonstrates robust performance across different distances and postures.

We further extend the framework to respiratory-pause scenarios. The derived linear relationship between estimated amplitude and breathing-time ratio enables a threshold-based pause-detection mechanism. Experimental results show that the proposed method achieves 88\% accuracy under pause conditions and reduces the MAE by 1.33 bpm compared with baseline methods. These results demonstrate that the proposed framework improves both regular breathing estimation and apnea-related pause detection, thereby advancing contact-free respiratory monitoring for IoT-enabled health applications.

\bibliographystyle{IEEEtran} 
\bibliography{IEEEabrv,references} 

\end{document}


\setlength{\columnsep}{0.245 in}

\appendices
\section{Proof of the Group Error Behaviors}
\label{appendix:proofA}

\renewcommand{\theequation}{A.\arabic{equation}}
\setcounter{equation}{0}

\renewcommand{\thefigure}{A.\arabic{figure}}
\setcounter{figure}{0}

\renewcommand{\thetable}{A.\arabic{table}}
\setcounter{table}{0}

\subsection*{A.1 Effects of Baseline Removal}

\textit{Proposition 3:} The group-dependent baseline $b_m$ is eliminated through group-wise normalization.

\begin{proof}
 From the definitions:
\begin{equation}
\begin{aligned}
\alpha_m &= \min_{n} x_{m,n} = \min_{n} (s_{m,n} + b_m + \epsilon_{m,n}) \\
&= b_m + \min_{n} (s_{m,n} + \epsilon_{m,n}) \\
y_{m,n} &= \frac{x_{m,n} - \alpha_m}{\beta_m} \\
&= \frac{s_{m,n} + \bcancel{b_m} + \epsilon_{m,n} - [\bcancel{b_m} + \min_{j} (s_{m,j} + \epsilon_{m,j})]}{\beta_m} \\
&= \frac{s_{m,n} - \min_{j} s_{m,j} + \epsilon_{m,n} - \min_{j} \epsilon_{m,j}}{\beta_m}
\end{aligned}
\end{equation}

The subtraction of $\alpha_m$ produces three fundamental effects that collectively explain the superiority of group-wise normalization:

\begin{enumerate}
    \item {Baseline elimination}: The group-dependent baseline $b_m$ is removed, ensuring robustness to environmental variations between measurement groups. This reduces interference from static environmental factors without affecting the breathing-induced dynamic components.
    
    \item {Signal relativization}: The term $s_{m,n} - \min_j s_{m,j}$ transforms absolute amplitudes into relative values by removing the group-wise bias $\min_j s_{m,j}$. This preserves the essential periodic structure of breathing signals while enabling direct comparison across subcarriers.
    
    \item {Noise impact}: The operation $\epsilon_{m,n} - \min_j \epsilon_{m,j}$ may alter the noise distribution, but when the respiratory signal dominates in (A.1), the impact on breathing frequency estimation is limited.
\end{enumerate} \end{proof}

\textit{Remark:}
Within each group $m$, $\alpha_m$ and $\beta_m$ are constants, which makes the normalization an affine transformation for all samples in that group. Although $\beta_m$ may vary across groups, the normalized signals retain the dominant respiratory patterns required for subsequent clustering and breathing estimation, as illustrated in Fig.~5. 

\subsection*{A.2 Comparison with Global Min-Max Normalization}

For global min-max normalization, the denominator is computed over all groups and all subcarriers:
\[
D_{\text{global}} =
\max_{m,j} x_{m,j} - \min_{m,j} x_{m,j}.
\]

For group-wise normalization, the denominator at group \(m\) is:
\[
\beta_m =
\max_j x_{m,j} - \min_j x_{m,j}.
\]

Since the global maximum and minimum are taken over the entire observation matrix, the global range satisfies
\[
D_{\text{global}} \geq \beta_m,\quad \forall m.
\]

Therefore, global normalization uses a larger (or equal) denominator whose value is determined by the entire observation matrix. As a result, extreme values and large-amplitude interference exert a greater influence on the scaling process, which tends to compress respiration-induced fluctuations.

\begin{figure}[t]
    \centering
    \includegraphics[width=0.65\linewidth]{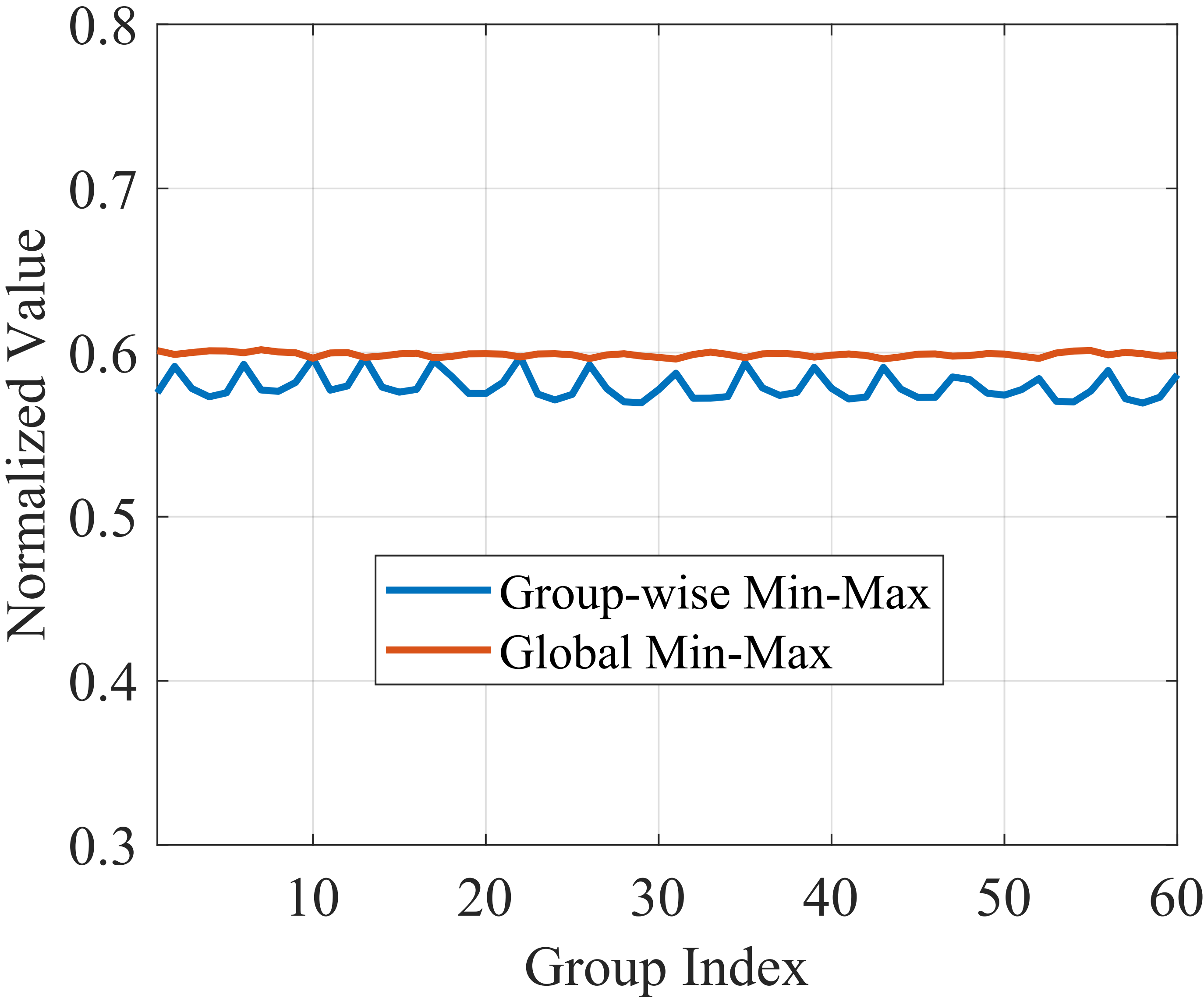}
    \caption{Comparison of global and group normalization schemes.}
    \label{fig:appendix1}
\end{figure}

Moreover, we compare the effects of the two normalization schemes. As shown in Fig. \ref{fig:appendix1}, for the sensitive subcarrier, global normalization produces substantially smaller amplitude fluctuations than group-wise normalization, which is consistent with the above analysis. These observations support the use of group-wise normalization in the proposed framework.

\section{Normalized Feature Relationships} 
\renewcommand{\theequation}{B.\arabic{equation}}
\setcounter{equation}{0}

\renewcommand{\thefigure}{B.\arabic{figure}}
\setcounter{figure}{0}

\renewcommand{\thetable}{B.\arabic{table}}
\setcounter{table}{0}

\subsection*{B.1 Statistical Feature Relationships}

\textit{Proposition 4:} For breathing-sensitive subcarriers with dominant sinusoidal breathing modulation of amplitude $\kappa_n$, the group-wise normalized features satisfy:
\begin{align}
\tilde{R}_n &\propto \tilde{\sigma}_n  \\
\tilde{E}_n &\propto (\tilde{\sigma}_n)^2 
\end{align}
where $\tilde{\sigma}_n$, $\tilde{R}_n$, and $\tilde{E}_n$ are the min-max normalized versions of $\sigma_n$, $R_n$, and $E_n$ respectively.

\begin{proof} To facilitate theoretical analysis, the following simplifying assumptions are made [6] [21].

Assumptions: (i) for breathing-sensitive subcarriers, the breathing modulation amplitude $\kappa_n$ dominates over noise ($\kappa_n \gg \sigma_\zeta$);
(ii) for insensitive subcarriers, the noise term $\delta_{m,i}$ is much smaller than the baseline $C_i$ ($|\delta_{m,i}| \ll C_i$);
(iii) noise variance is consistent across subcarriers.

After normalization to $[0,1]$ range:
\begin{equation}
\tilde{\sigma}_n = \frac{\sigma_n - \sigma_{\min}}{\sigma_{\max} - \sigma_{\min}}, \quad
\tilde{R}_n = \frac{R_n - R_{\min}}{R_{\max} - R_{\min}}, \quad
\tilde{E}_n = \frac{E_n - E_{\min}}{E_{\max} - E_{\min}}.
\end{equation}
where $\sigma_{\min}$, $R_{\min}$, $E_{\min}$ and $\sigma_{\max}$, $R_{\max}$, $E_{\max}$ are the minimum and maximum values, respectively.

For breathing-sensitive subcarriers with raw features (neglecting noise contribution):
\begin{align}
\sigma_n = \frac{\kappa_n}{\sqrt{2}}, \ R_n = 2\kappa_n, \
E_n = \frac{M\kappa_n^2}{2}.
\end{align}

The normalization constants are determined by extreme values across all subcarriers:
\begin{align}
\sigma_{\max} &= \frac{\kappa_{\max}}{\sqrt{2}}, \quad R_{\max} = 2\kappa_{\max}, \quad E_{\max} = \frac{M\kappa_{\max}^2}{2}, \nonumber \\
\
\sigma_{\min} &\approx 0, \quad R_{\min} \approx 0, \quad E_{\min} \approx 0.
\end{align}

Thus, for a sensitive subcarrier with amplitude $\kappa_n$:
\begin{align}
\tilde{\sigma}_n &= \frac{\kappa_n}{\kappa_{\max}}, \
\tilde{R}_n = \frac{\kappa_n}{\kappa_{\max}}, \
\tilde{E}_n = \left(\frac{\kappa_n}{\kappa_{\max}}\right)^2.
\end{align}

This yields the normalized relationship:
\begin{equation}
\tilde{R}_n \propto \tilde{\sigma}_n, \quad \tilde{E}_n \propto (\tilde{\sigma}_n)^2.
\end{equation}

\noindent\textbf{For breathing-insensitive subcarriers:}

These subcarriers contain negligible breathing modulation, and their normalized signal can be modeled as:
\begin{equation}
y_n(m) = \mu_n + \zeta_n(m)
\end{equation}
where $\zeta_n(m)$ is zero-mean noise with variance $\sigma_\zeta^2$. The baseline $\mu_n$ does not affect the standard deviation (STD) or range, as they are computed after mean subtraction; both $\sigma_n$ and $R_n$ are therefore determined solely by the noise term.

After min-max normalization, the feature values of all subcarriers are mapped to the $[0,1]$ interval. For breathing-insensitive subcarriers, their raw features arise primarily from noise and are typically smaller than those of sensitive subcarriers. Thus, breathing-insensitive subcarriers cluster near the origin in the normalized feature space, while sensitive subcarriers concentrate in the regions of larger feature values along $\tilde{R} \propto \tilde{\sigma}$ and $\tilde{E} \propto \tilde{\sigma}^2$.
\end{proof}

\subsection*{B.2 Impact of Motions on Feature Relationships}
\begin{figure}[t]
    \centering
    \includegraphics[width=0.65\linewidth]{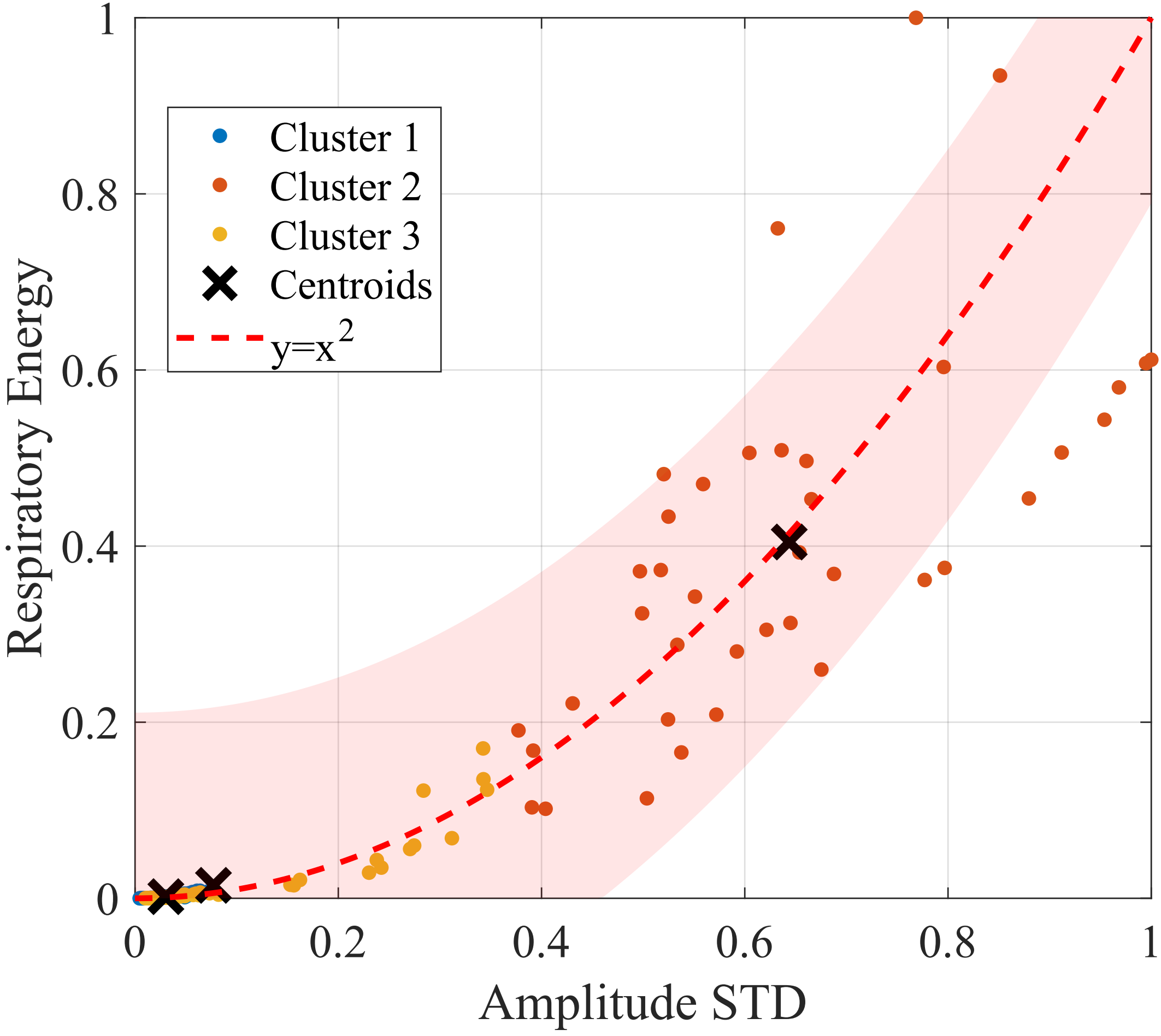}
    \caption{ Illustration of feature relationships under motion interferences.}
    \label{fig:appendix2}
\end{figure}

To examine the impact of the derived feature relationships under non-stationary conditions, we conducted an additional experiment with motion interference (a walking person in the room). As shown in Fig.~\ref{fig:appendix2}, most subcarriers remain close to the quadratic trend, which indicates that the theoretical relationship is qualitatively robust. However, the 90\% coverage band (shaded region) becomes notably wider than in the quiet case (Fig. 6), which indicates that motion interference introduces additional variance. Despite this broadening, sensitive subcarriers remain separable from insensitive ones. This analysis suggests that while the proposed feature relationships hold under moderate disturbances, more sophisticated models may be needed for highly dynamic environments, as discussed in Section~V-G.

\section{Theoretical Analysis of Breathing Pauses on Sinusoidal Fitting}
\renewcommand{\theequation}{C.\arabic{equation}}
\setcounter{equation}{0}

\renewcommand{\thefigure}{C.\arabic{figure}}
\setcounter{figure}{0}

\renewcommand{\thetable}{C.\arabic{table}}
\setcounter{table}{0}

\subsection*{C.1 Amplitude Attenuation}

\textit{Proposition 5:} When fitting a sinusoidal model to a signal containing both breathing and pause periods, the estimated amplitude $\hat{A}$ exhibits an approximately linear relationship with the breathing time ratio $T_b/T$, with oscillations due to the non‑integer‑cycle truncation:

\begin{equation}
\frac{\hat{A}}{A} \approx \frac{T_b}{T} + \delta\left(\cdot \right)
\end{equation}
where $\delta(\cdot)$ is an oscillatory correction term.

\begin{proof}
To derive the optimal amplitude estimate $\hat{A}$, we minimize the objective function $J$ with respect to $\hat{A}$. The derivative of $J$ with respect to $\hat{A}$ is:

\begin{align}
    \frac{\partial J}{\partial \hat{A}} &= \frac{\partial}{\partial \hat{A}} \left[ \int_0^{T_b} \left( s(t) - \hat{A} \sin(2\pi \hat{r_b} t + \hat{\phi}) - \hat{C} \right)^2 dt \right. \nonumber \\
    &\quad + \left. \int_{T_b}^T \left( \hat{A} \sin(2\pi \hat{r_b} t + \hat{\phi}) + \hat{C} \right)^2 dt \right] \nonumber \\
    &= -2 \int_0^{T_b} \left( s(t) - \hat{A} \sin(2\pi \hat{r_b} t + \hat{\phi}) - \hat{C} \right) \sin(2\pi \hat{r_b} t + \hat{\phi}) dt \nonumber \\
    &\quad + 2 \int_{T_b}^T \left( \hat{A} \sin(2\pi \hat{r_b} t + \hat{\phi}) + \hat{C} \right) \sin(2\pi \hat{r_b} t + \hat{\phi}) dt
    \label{eq:derivative_A}
\end{align}

Setting $\partial J / \partial \hat{A} = 0$ for optimality and assuming $\hat{C} \approx 0$ (which holds as the baseline is normalized), we obtain:

\begin{equation}
\int_0^{T_b} s(t) \sin(2\pi \hat{r_b} t + \hat{\phi}) dt = \hat{A} \int_0^{T} \sin^2(2\pi \hat{r_b} t + \hat{\phi}) dt
\label{eq:optimal_amplitude}
\end{equation}

Assuming the estimated parameters approach the true values ($\hat{r_b} \approx r_b$, $\hat{\phi} \approx \phi$) and $\hat{C} \approx 0$, the optimal amplitude estimate from~\eqref{eq:optimal_amplitude} can be expressed as:

\begin{equation}
\hat{A} = A \cdot \frac{\int_0^{T_b} \sin^2(2\pi r_b t + \phi) dt}{\int_0^T \sin^2(2\pi r_b t + \phi) dt}
\label{eq:amplitude_ratio_general}
\end{equation}

Evaluating these integrals explicitly:

\begin{align}
&\int_0^{T_x} \sin^2(2\pi r_b t + \phi) dt \nonumber \\
&\quad = \frac{T_x}{2} - \frac{1}{8\pi r_b} \left[ \sin(4\pi r_b T_x + 2\phi) - \sin(2\phi) \right] \nonumber \\
&\quad = \frac{T_x}{2} \left[ 1 - \frac{\sin(4\pi r_b T_x + 2\phi) - \sin(2\phi)}{4\pi r_b T_x} \right]
\label{eq:sin2_integral_form}
\end{align}

Substituting into (\ref{eq:amplitude_ratio_general}):

\begin{align}
\frac{\hat{A}}{A} &= \frac{T_b}{T} \cdot 
\frac{1 - \frac{\sin(4\pi r_b T_b + 2\phi) - \sin(2\phi)}{4\pi r_b T_b}}
{1 - \frac{\sin(4\pi r_b T + 2\phi) - \sin(2\phi)}{4\pi r_b T}} \nonumber \\
&= \frac{T_b}{T} \cdot \frac{1 - \epsilon_b}{1 - \epsilon_T}
\label{eq:amplitude_ratio_with_epsilon}
\end{align}
where $\epsilon_x = \frac{\sin(4\pi r_b T_x + 2\phi) - \sin(2\phi)}{4\pi r_b T_x}$ for $x \in \{b, T\}$.

\subsubsection{Linear Approximation with Oscillation Correction}

Using Taylor expansion for $|\epsilon_T| \ll 1$, \eqref{eq:amplitude_ratio_with_epsilon} becomes:

\begin{align}
\frac{\hat{A}}{A} &= \frac{T_b}{T} \cdot (1 - \epsilon_b) \cdot (1 + \epsilon_T + \epsilon_T^2 + \cdots) \nonumber \\
&\approx \frac{T_b}{T} \left[ 1 + (\epsilon_T - \epsilon_b) + \mathcal{O}(\epsilon^2) \right] 
\label{eq:amplitude_ratio_approx}
\end{align}

The dominant behavior is the linear term $T_b/T$, which provides a simple approximation for the amplitude ratio:

\begin{equation}
\hat{A} \approx A \cdot \frac{T_b}{T}
\label{eq:amplitude_approx}
\end{equation}

The accuracy of this approximation is determined by the oscillatory correction term $(\epsilon_T - \epsilon_b)$. Since each numerator satisfies $|\sin(\cdot)-\sin(2\phi)| \leq 2$, we have $|\epsilon_T| \leq 1/(2\pi r_b T)$ and $|\epsilon_b| \leq 1/(2\pi f T_b)$. Therefore, the approximation error is bounded by:

\begin{equation}
\left| \frac{\hat{A}}{A} - \frac{T_b}{T} \right| \leq \frac{T_b}{T} \left( \frac{1}{2\pi r_b T} + \frac{1}{2\pi r_b T_b} \right) \leq \frac{1}{\pi r_b T}
\label{eq:oscillation_amplitude}
\end{equation}
where the last inequality uses $T_b \leq T$. This bound shows that the linear approximation becomes increasingly accurate as the observation window length $T$ increases.

\subsubsection{Special Case Analysis}

When $T_b = n/r_b$ (integer number of respiratory cycles), the numerator simplifies exactly, yielding:

\begin{equation}
\frac{\hat{A}}{A} = \frac{n/r_b}{T} \cdot \frac{1}{1 - \epsilon_T} \approx \frac{T_b}{T} \quad \text{for } r_bT \gg 1.
\label{eq:integer_cycles_correct}
\end{equation}

The estimation error is minimized, demonstrating the importance of analyzing signal segments containing integer respiratory cycles.

When $T_b \rightarrow 0$, the signal approaches a pure pause. The least-squares solution converges to:

\begin{align}
    \hat{A} &\rightarrow 0 \\
    \hat{r_b} &\rightarrow \text{undefined (fitting unstable)} \\
    \hat{C} &\rightarrow 0
    \label{eq:pure_apnea}
\end{align}

This highlights the necessity of sufficient breathing duration for reliable parameter estimation.

Overall, this analysis derives the approximate linear attenuation $\hat{A}/A \approx T_b/T$ with oscillations. This quantifies the amplitude suppression induced by pause periods, forming the theoretical basis for the threshold-based pause detection method in Section IV-D.
\end{proof}

\subsubsection{Simulation Observation}
We further examined amplitude estimation under pause conditions:
\begin{itemize}
    \item {Signal model:} $s(t)=A\sin(2\pi r_b t+\phi)+\eta(t)$ with $\eta(t)\sim\mathcal{N}(0,0.05^2)$.
    \item {Window lengths:} $T\in\{3,5,10\}$ s with $T_b/T=0.5$ for mixed case.
\end{itemize}

Results in Fig.9 confirm the linear compression trend $\hat{A}\approx A\cdot(T_b/T)$ as the leading-order behavior, with oscillatory deviations that decrease as $1/T$.

\subsubsection{Real CSI Observation}

\begin{figure}[t]
\centering
\includegraphics[width=0.65\linewidth]{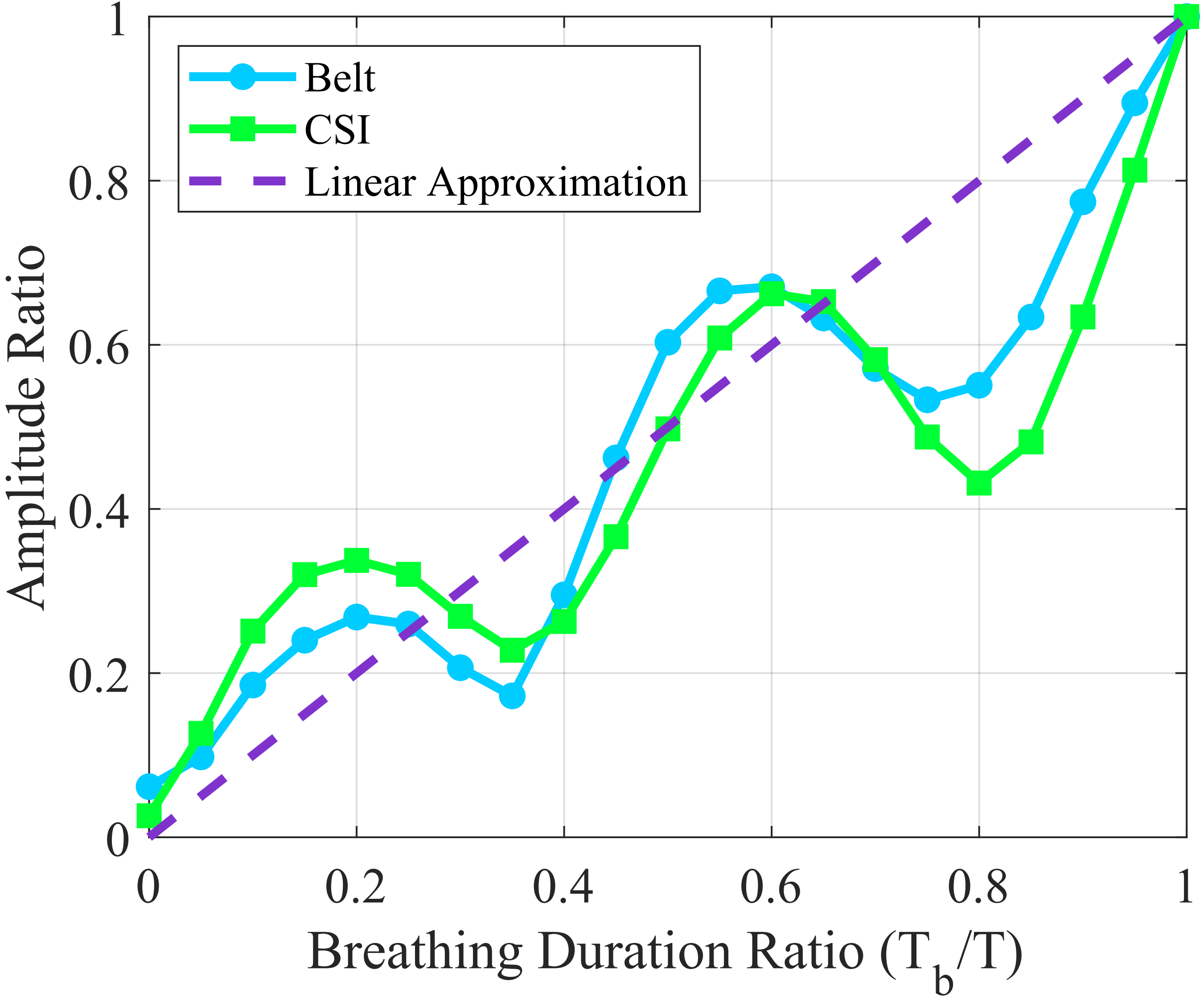}
\caption{Observations in real CSI and respiration-belt signals.}
\label{fig:stop4}
\end{figure}

To further support the theoretical analysis, we also examined amplitude attenuation using both respiration-belt and real CSI signals collected during breathing pauses. As shown in Fig.~\ref{fig:stop4}, the normalized fitted amplitudes approximately follow the linear attenuation trend while exhibiting oscillatory deviations. These observations are consistent with the theoretical and simulation results.

\subsection*{C.2 Frequency Estimation}
\addcontentsline{toc}{subsection}{C.2 Another Subsection}
\setcounter{subsubsection}{0} 

\textit{Proposition 6: }
When fitting a sinusoidal model to a respiratory signal containing a pause interval, the breathing frequency estimate $\hat{r_b}$ is predominantly determined by the breathing segment, while the pause segment theoretically contributes no frequency-selective information and therefore introduces only limited bias.

\begin{proof}
The frequency estimation bias can be analyzed based on the optimization condition derived from the least-squares cost function:

\begin{align}
\frac{\partial J}{\partial \hat{f}_b} &= -4\pi \hat{A} \int_0^{T_b} \left( s(t) - \hat{s}(t) \right) t \cos(2\pi \hat{r_b} t + \hat{\phi}) \, dt \nonumber \\
&\quad + 4\pi \hat{A} \int_{T_b}^{T} \hat{s}(t) \, t \cos(2\pi \hat{r_b} t + \hat{\phi}) \, dt = 0
\label{eq:opt_cond}
\end{align}

This condition reveals fundamentally different roles of the breathing and pause segments in frequency estimation.

\subsubsection{Pause Segment: Frequency-Independent Constraint}

During pause ($T_b \leq t \leq T$), the true signal satisfies $s(t) \approx 0$. Consider fitting the pause segment with a sinusoidal model:

\begin{equation}
\varepsilon_{\text{pause}}(\hat{r_b}) = \int_{T_b}^T \left[ \hat{A} \sin(2\pi\hat{r_b}t + \hat{\phi}) \right]^2 dt
\end{equation}

For any fixed $\hat{r_b}$, this error function can be minimized by setting $\hat{A}=0$, yielding the minimum error:

\begin{equation}
\min_{\hat{A},\hat{\phi}} \varepsilon_{\text{pause}}(\hat{r_b}, \hat{A}, \hat{\phi}) = 0 \quad \text{for \textit{any} frequency } \hat{r_b}.
\end{equation}

This demonstrates that the pause segment provides no frequency-selective information; any frequency can fit the zero signal equally well when the amplitude is optimized. Consequently, the second term in \eqref{eq:opt_cond} contributes only a weak, frequency-flat constraint to the overall optimization landscape.

\subsubsection{Breathing Segment: Frequency-Selective Constraint}

During effective breathing ($0 \leq t < T_b$), the true signal is $s(t) = A\sin(2\pi r_b t + \phi)$. For a candidate frequency $\hat{r_b}$, assuming optimal amplitude and phase alignment ($\hat{A}=A$, $\hat{\phi}=\phi$), the residual becomes:

\begin{equation}
s(t) - \hat{s}(t) = A\left[\sin(2\pi r_b t) - \sin(2\pi\hat{r_b}t)\right]
\end{equation}

Using the sum-to-product identity:
\begin{equation}
\sin u - \sin v = 2 \cos\left(\frac{u+v}{2}\right) \sin\left(\frac{u-v}{2}\right)
\end{equation}
where $u = 2\pi r_b t$ and $v = 2\pi \hat{r_b} t$, we obtain:

\begin{equation}
s(t) - \hat{s}(t) = 2A \cos\left(2\pi\frac{r_b+\hat{r_b}}{2}t\right) \sin\left(2\pi\frac{r_b-\hat{r_b}}{2}t\right).
\label{eq:residual_beat}
\end{equation}

This reveals a beat frequency at $\Delta r_b = |r_b - \hat{r_b}|$. The integrated squared error over the breathing segment is:

\begin{equation}
\varepsilon_{\text{breath}}(\Delta r_b) = \int_0^{T_b} \left[ s(t) - \hat{s}(t) \right]^2 dt
\end{equation}

This error term is minimized when $\hat{r_b} = r_b$ and grows as the candidate frequency deviates from the true breathing frequency. In contrast to the breathing pause segment, this frequency-selective behavior plays a dominant role in frequency estimation of breath signals. Furthermore, the beat frequency structure causes $\varepsilon_{\text{breath}}(\Delta r_b)$ to oscillate with $\hat{r_b}$, with local minima corresponding to integer multiples of $1/T_b$ away from $r_b$.
\end{proof}

\subsubsection{Simulation Validation}

To validate Proposition~6 regarding the impact of breathing pauses on sinusoidal fitting, we conducted numerical simulations to examine these effects. The simulation setup follows:
\begin{itemize}
    \item {Breathing signal:} $s(t)=A\sin(2\pi r_b t+\phi)$ with $A=1$, $r_b=0.3$ Hz, $\phi=0$, sampling rate $r=100$ Hz.
    \item {Observation window:} The total signal comprises a breathing segment of $T_b$ followed by a fixed pause segment of $T_a=5$ s, where $s(t)=0$. The total window length is $T_{total}=T_b+T_a$, and we vary $T_b$ to examine its impact on estimation accuracy.
    \item {Frequency scan:} Candidate $\hat{r_b}\in[0.1,0.5]$ Hz with step $0.01$ Hz.
    \item {Error metric:} Average least-squares error per unit time $J_{\text{avg}}(\hat{r_b}) = \frac{1}{N} \sum_{t=1}^{N} [s(t) - \hat{s}(t)]^2$, where $N = r \cdot T_{\text{total}}$ is the total number of samples.

\end{itemize}
As shown in Fig.~10, the average error per unit time consistently exhibits a pronounced minimum at the true breathing frequency ($r_b = 0.3$ Hz) across different observation window lengths. This confirms that the breathing segment dominates the frequency estimation, as the pause segment contributes no frequency-selective information. The oscillatory structure of the error curves also reflects the beat frequency effect derived in~\eqref{eq:residual_beat}, with local minima corresponding to frequency deviations that are integer multiples of $1/T_b$. Furthermore, the average error per unit time increases with the total observation window $T_{\text{total}}$. This is because longer windows contain a larger proportion of the breathing signal, which amplifies the error for frequencies deviating from the true breathing rate. This observation motivates our use of a shorter time window for breath estimation during pause episodes in Section~IV-D.

\vfill